\documentclass[lettersize,journal]{IEEEtran}
\usepackage{comment}
\usepackage{xcolor}
\usepackage{amsmath,amsfonts}
\usepackage{algorithmic}
\usepackage{algorithm}
\usepackage{array}
\usepackage[caption=false,font=normalsize,labelfont=sf,textfont=sf]{subfig}
\usepackage{textcomp}
\usepackage{stfloats}
\usepackage{url}
\usepackage{verbatim}
\usepackage{graphicx}
\usepackage{cite}
\usepackage{tabularx}
\usepackage{booktabs}
\usepackage{longtable}
\usepackage{enumitem}
\usepackage{etoolbox}
\usepackage{makecell, multirow, colortbl}

\ifCLASSINFOpdf \else \fi
\IEEEoverridecommandlockouts

\begin{document}

\title{Zero-Trust Foundation Models: A New Paradigm for Secure and Collaborative Artificial Intelligence for Internet of Things}

\author{Kai Li, \IEEEmembership{Senior Member, IEEE}, Conggai Li, Xin Yuan,  \IEEEmembership{Senior Member, IEEE}, Shenghong Li,  \IEEEmembership{Member, IEEE},\\
Sai Zou, \IEEEmembership{Senior Member, IEEE}, 
Syed Sohail Ahmed, 
Wei Ni, \IEEEmembership{Fellow, IEEE}, Dusit Niyato, \IEEEmembership{Fellow, IEEE}, \\
Abbas Jamalipour, \IEEEmembership{Fellow, IEEE}, Falko Dressler, \IEEEmembership{Fellow, IEEE},  
and Özgür B. Akan, \IEEEmembership{Fellow, IEEE}.
\thanks{K.~Li and F.~Dressler are with the School of Electrical Engineering and Computer Science, TU Berlin, Germany. K. Li is also with Real-Time and Embedded Computing Systems Research Centre (CISTER), Porto 4249--015, Portugal (e-mail: kaili@ieee.org, dressler@ccs-labs.org).

C. Li, X. Yuan, S. Li, and W. Ni are with Data61, CSIRO, Sydney, NSW 2122, Australia. X. Yuan and W. Ni are also with the School of Computer Science and Engineering, the University of New South Wales, Kensington, NSW 2033, Australia (e-mail: \{conggai.li, xin.yuan, shenghong.li, wei.ni\}@csiro.au).

S. Zou is with the State Key Laboratory of Public Big Data, College of Big Data and Information Engineering, Guizhou University, Guiyang 550025, China (e-mail: dr-zousai@foxmail.com).

S. S. Ahmed is with the Department of Computer Engineering, Qassim University, Buraydah 52571, Kingdom of Saudi Arabia (e-mail: sa.ahmed@qu.edu.sa).

D. Niyato is with the College of Computing and Data Science, Nanyang Technological University, Singapore 639798, Singapore (e-mail: dniyato@ntu.edu.sg).

A. Jamalipour is with the School of Electrical and Information Engineering, the University of Sydney, Sydney, NSW 2006, Australia (e-mail: a.jamalipour@ieee.org).


O. B. Akan is with the Division of Electrical Engineering, Department of Engineering, University of Cambridge, CB3 0FA Cambridge, U.K., and also with the Center for NeXt-Generation Communications (CXC), Ko\c c University, 34450 Istanbul, Turkey (e-mail: oba21@cam.ac.uk).

}

}



\maketitle

\begin{abstract}
This paper focuses on Zero-Trust Foundation Models (ZTFMs), a novel paradigm that embeds zero-trust security principles into the lifecycle of foundation models (FMs) for Internet of Things (IoT) systems. By integrating core tenets, such as continuous verification, least privilege access (LPA), data confidentiality, and behavioral analytics into the design, training, and deployment of FMs, ZTFMs can enable secure, privacy-preserving AI across distributed, heterogeneous, and potentially adversarial IoT environments. We present the first structured synthesis of ZTFMs, identifying their potential to transform conventional trust-based IoT architectures into resilient, self-defending ecosystems. Moreover, we propose a comprehensive technical framework, incorporating federated learning (FL), blockchain-based identity management, micro-segmentation, and trusted execution environments (TEEs) to support decentralized, verifiable intelligence at the network edge. In addition, we investigate emerging security threats unique to ZTFM-enabled systems and evaluate countermeasures, such as anomaly detection, adversarial training, and secure aggregation. Through this analysis, we highlight key open research challenges in terms of scalability, secure orchestration, interpretable threat attribution, and dynamic trust calibration. This survey lays a foundational roadmap for secure, intelligent, and trustworthy IoT infrastructures powered by FMs.
\end{abstract}


\IEEEpeerreviewmaketitle





\begin{table}[htb]
\centering
\caption{Abbreviation and full name}
\begin{tabularx}{8cm}{l|X}
\hline
\bf{Abbreviation} & \bf{Full name} \\ \hline
AI & Artificial Intelligence \\ 
CPS & Cyber-Physical Systems \\
FedIoT & Federated Learning-enabled IoT \\ 
FL & Federated Learning \\
FM & Foundation Models \\
GNN & Graph Neural Network \\
IDS & Intrusion Detection Systems \\
IoT & Internet of Things \\ 
IIoT & Industrial IoT \\
JIT & Just-In-Time \\
LLMs & Large Language Models \\
LPA & Least Privilege Access \\
LSTM & Long Short-Term Memory \\
ML & Machine Learning \\
NLP & Natural Language Processing \\
SMPC & Secure Multi-Party Computation \\
SOTA & State-Of-The-Art \\
TEEs & Trusted Execution Environments \\
ZTFM & Zero-Trust Foundation Models \\
\hline
\end{tabularx}
\label{tb_names}
\end{table}

\section{Introduction}
\subsection{Artificial Intelligence for Internet of Things}

\IEEEPARstart{A}{rtificial} Intelligence (AI) is driving a paradigm shift in the Internet of Things (IoT), enabling intelligent, data-driven decision-making across distributed, sensor-rich environments~\cite{hopkins2021internet,hu2021distributed}. By enhancing the perception, reasoning, and actuation capabilities of IoT devices, AI facilitates smarter automation and system-wide optimization in sectors such as manufacturing~\cite{trakadas2020artificial}, agriculture~\cite{singh2023blockchain}, transportation~\cite{munir2020drive}, and home automation~\cite{reena2018intelligent}.

According to the ``Digital Transformation Enabler: Machine Learning'' report from the Industry IoT Consortium~\cite{iiconsortium2025}, AI-enabled IoT systems already deliver tangible benefits. In manufacturing, AI-driven predictive maintenance reduces equipment downtime and boosts operational efficiency. For instance, KONUX applies AI-powered sensors to monitor railway infrastructure in real time~\cite{konux2025}, enhancing public safety and reliability. In agriculture, precision farming leverages AI algorithms to analyze soil conditions, forecast weather, and optimize irrigation. Consumer platforms like LG’s ThinQ ON hub~\cite{LG2025} use machine learning (ML) to manage smart home devices based on user behavior and contextual patterns.

Building upon these domain-specific applications, a new class of models -- Foundation Models (FMs) -- has emerged as the next frontier in AI. Unlike task-specific models, FMs are pre-trained on massive datasets and exhibit cross-domain generalization capabilities through self-supervised learning~\cite{yang2025survey}. Notable examples, e.g., OpenAI's GPT~\cite{Brown2020nips}, Google's BERT~\cite{Devlin2019naacl}, and China's DeepSeek~\cite{abs-2501-12948corr, abs-2412-19437corr}, support a wide range of downstream applications, from natural language understanding to multimodal reasoning. Their integration into IoT offers new opportunities for decentralized intelligence, adaptive control, and autonomous collaboration. 

\subsection{Security Challenges of Foundation Models}

The scale and complexity of FMs present critical challenges when deployed in resource-constrained and adversarial IoT environments. Traditional AI pipelines fall short in addressing key issues such as data confidentiality, trust enforcement, energy efficiency, and security assurance. These limitations can be prominent in IoT systems, where devices are heterogeneous, intermittently connected, and often physically exposed.

Despite their expressiveness capability, FMs bring significant security and privacy risks in IoT deployments. Unlike traditional ML models, FMs are typically pre-trained on massive, heterogeneous datasets and deployed across distributed infrastructures~\cite{boateng2025survey}. This decentralized nature disrupts conventional security assumptions and expands the attack surface, leaving systems vulnerable to membership inference~\cite{mattern2023membership}, model poisoning, backdoor injections, and adversarial inference attacks~\cite{li2023backdoor}.

Energy consumption emerges as a key performance metric. FMs demand substantial computational resources, which poses challenges for energy-constrained IoT devices. An effective FM framework is expected to strike a balance between robust security, model accuracy, and energy efficiency.

Other crucial metrics include data integrity, communication overhead, and latency, all of which are key to many mission-critical IoT scenarios such as smart healthcare, industrial automation, and autonomous vehicles. The diversity and sensitivity of IoT-generated data (e.g., medical records or operational telemetry) require scalable, lightweight privacy-preserving techniques, e.g., differential privacy and secure federated aggregation, that go beyond what is used in centralized FM infrastructures~\cite{yu2023federated}.

These considerations substantiate the urgent need for a paradigm shift towards a unified framework that integrates the expressiveness of FMs with the security, privacy, and efficiency guarantees of a zero-trust architecture, tailored specifically for the constraints and requirements of IoT ecosystems.

\begin{figure}
    \centering
    \includegraphics[width=0.9\linewidth]{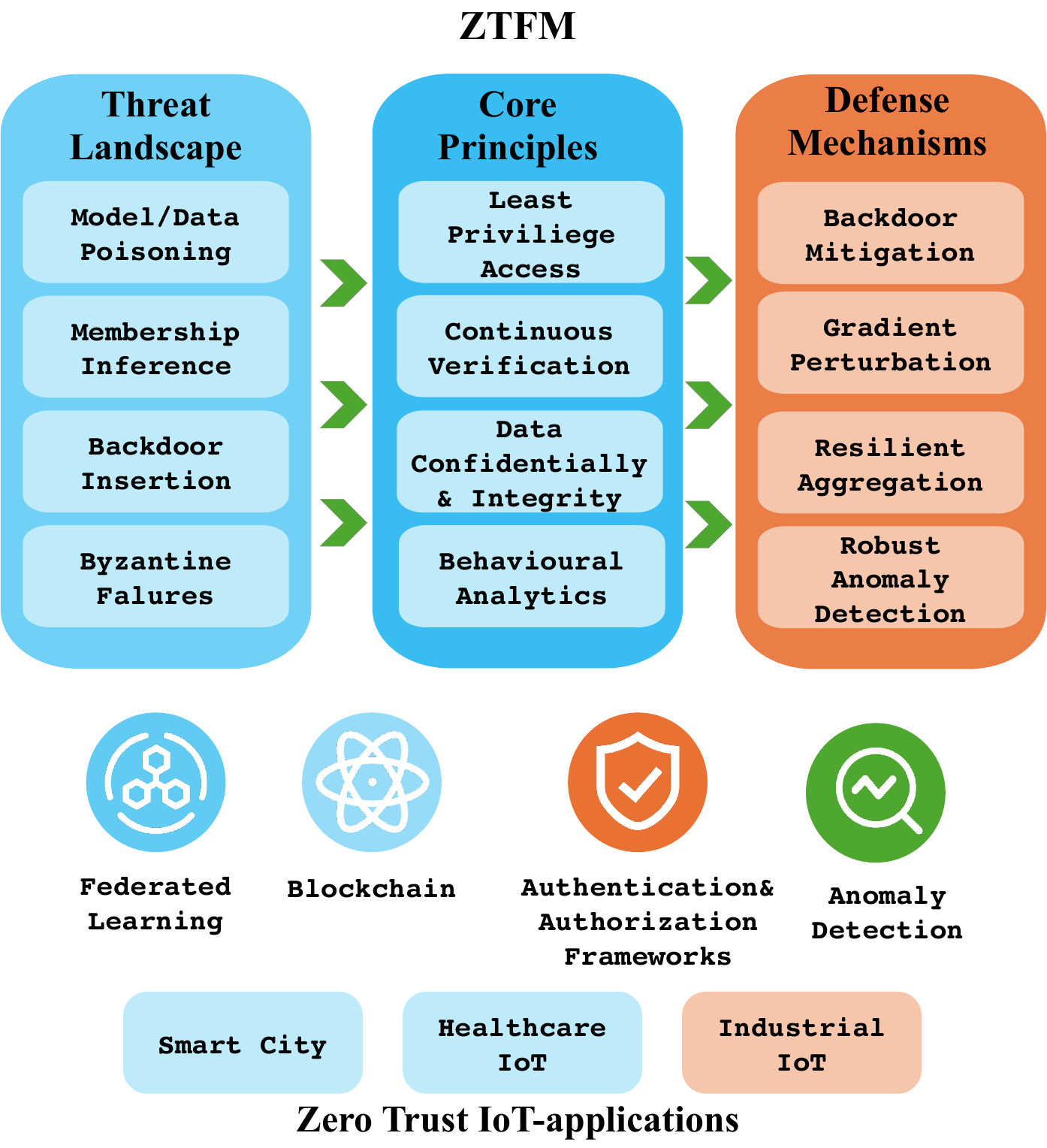}
    \caption{{\color{black}A conceptual overview of ZTFM-enabled IoT applications. The architecture integrates Zero Trust principles with enabling technologies. These components collaboratively mitigate adversarial threats such as model poisoning, membership inference, backdoors, and Byzantine failures, enhancing the security posture of real-world IoT scenarios, including Smart Cities, Healthcare, and Industrial IoT.}}
    \label{fig:applications}
\end{figure}

\subsection{Contributions}
This paper advocates for embedding zero-trust security principles~\cite{egerton2021applying} into the FM lifecycle-enforcing continuous verification, least privilege access (LPA), and secure computation by design, and puts forth a new zero-trust FM (ZTFM) for IoT systems and applications. 
A ZTFM is envisaged to integrate zero-trust security principles into the design, training, and deployment of FMs, enabling continuous verification, fine-grained access control, and privacy-preserving computation across distributed IoT environments. By integrating zero-trust security with the adaptive, intelligent capabilities of FMs, IoT systems can move from being vulnerable, trust-based networks to self-defending ecosystems. The result is a more secure, more autonomous IoT, which is able to overcome the challenges of a dynamic and increasingly hostile digital landscape.

This paper represents the first structured effort to define, formalize, and analyze ZTFM in the context of IoT. By examining how core zero-trust principles, including \textit{LPA}, \textit{Continuous Verification}, \textit{Data Confidentiality and Integrity}, and \textit{Behavioral Analytics}, can be operationalized in FM-driven IoT systems, we aim to bridge the gap between trustless IoT architectures and scalable AI frameworks.
We also provide a technical synthesis of key enabling technologies for ZTFMs, including federated learning (FL), blockchain-based identity management, micro-segmentation, and trusted execution environments (TEEs), outlining how they jointly support secure, collaborative AI across heterogeneous IoT and edge networks. Moreover, we analyze the unique threats and limitations that arise in ZTFM-enabled IoT environments and identify future research opportunities for scalable, interpretable, and resilient trust enforcement in real-world IoT applications. 

The contributions of this paper are summarized as follows:

\begin{enumerate}
    \item We present the first comprehensive synthesis of ZTFMs in IoT systems and applications, identifying a timely opportunity to integrate zero-trust principles with the training and deployment of FMs. This integration addresses the growing need for dynamic, decentralized trust management in AI-driven IoT systems.

    \item We formalize four core security principles and analyze their operationalization in constrained, 
    potentially malicious,
    and heterogeneous IoT environments. This
    helps reveal implementation gaps in current zero-trust deployments and highlights open research directions for principle-level enforcement in FM workflows.

    \item We propose a unified technical framework of ZTFMs that combines FL, blockchain-based identity management, micro-segmentation, and TEEs. This framework not only enables privacy-preserving and verifiable AI computation at the network edge, but also identifies current limitations in scalability, secure orchestration, and interoperability across IoT infrastructures.

    \item We conduct an in-depth analysis of emerging threats in ZTFM-enabled IoT systems and review defense strategies, such as anomaly detection, adversarial training, and secure aggregation. Accordingly, we identify future research challenges, including lightweight secure multiparty computation (SMPC), interpretable threat attribution, and AI-driven dynamic trust calibration for real-time edge intelligence in future intelligent IoT systems.
\end{enumerate}

\subsection{Promising Applications of ZTFMs for Internet of Things} 
\label{sectionII}

The combination of zero-trust principles and FMs is poised to transform the security of the IoT. In a traditional IoT environment, devices are often trusted by default once they are authenticated -- a risky assumption given the growing scale and sophistication of cyber threats~\cite{makhdoom2018anatomy}. A zero-trust approach, summarized by the maxim ``never trust, always verify," insists that no device, user, or system be trusted automatically~\cite{kindervag2010build}. Every interaction must be continuously authenticated, authorized, and validated.

FMs, which are large-scale AI models trained on vast and diverse data, bring a new level of intelligence to implementing zero-trust for IoT. They can learn complex patterns of behavior across different types of devices and contexts~\cite{bommasani2021opportunities,liu2024survey}. By ingesting telemetry, logs, sensor outputs, and network behavior, these models build a detailed, evolving understanding of what ``normal" looks like for each device and system~\cite{liu2022trustworthy}. This enables them to continuously verify the legitimacy of device actions, detecting subtle signs of compromise or malfunction without relying on pre-written signatures or manual rules.

In practice, a ZFTM for IoT would monitor device behavior at scale, dynamically adjusting access permissions based on ongoing risk assessments~\cite{syed2022zero}. If a smart thermostat begins sending large volumes of encrypted data at odd hours, the model could immediately recognize this deviation from normal behavior, isolate the device, and alert administrators. Access control becomes dynamic and context-aware, based not only on static identities but on live, real-world behavior.

ZFTM can also automate policy generation and enforcement. As new devices are introduced into the network, the FM could propose security policies tailored to the device’s expected behavior and risk profile, reducing administrative burden. Over time, the system would adapt to changes without human intervention, maintaining a strong security posture as IoT ecosystems evolve~\cite{mensah2024zero}.

With advances in edge AI and FL, lightweight versions of these models could be deployed on gateways or edge devices~\cite{merenda2020edge}. This ensures that zero-trust principles are enforced even when connectivity to a central cloud is limited, preserving resilience and privacy in sensitive environments like smart homes~\cite{ashibani2019design}, factories~\cite{xu2024design}, supply chains~\cite{Lulu2025IOTJ}, and healthcare facilities~\cite{alsuwaidi2024transformative}.
Beyond these terrestrial applications, Mao et al.~\cite{Mao2025JSAC} explored a blockchain-enabled cold-start aggregation scheme for federated reinforcement learning in zero-trust LEO satellite networks. Their approach demonstrates that ZTFM principles can be extended to highly dynamic and adversarial edge environments, such as space-based IoT systems, enabling secure and privacy-preserving model updates even in scenarios with intermittent connectivity and untrusted participants.

\begin{figure}
    \centering
    \includegraphics[width=1\linewidth]{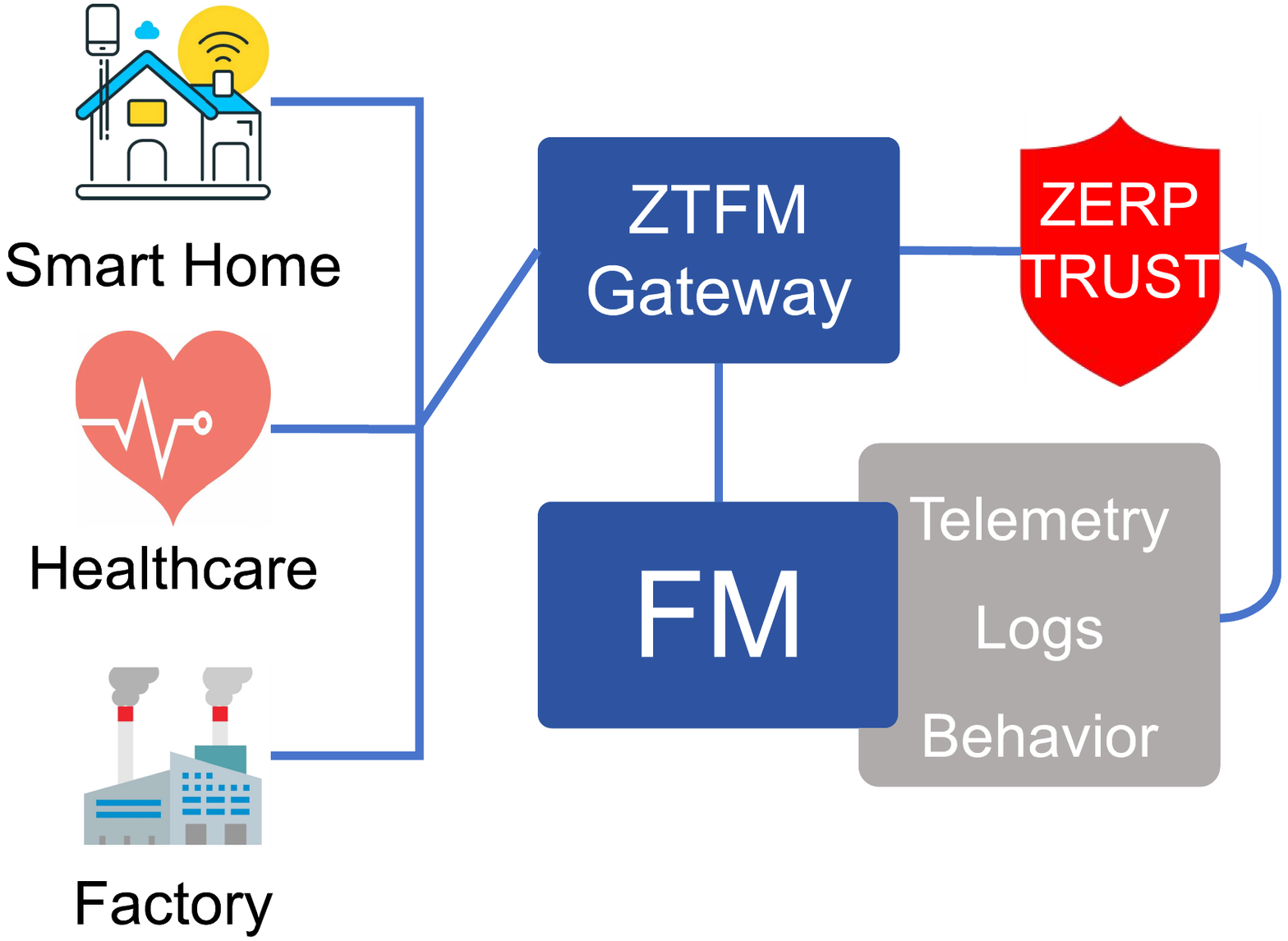}
    \caption{ZTFM-enabled security architecture for IoT. FMs analyze behavioral data to enforce Zero-Trust decisions and trigger alerts across smart home, healthcare, and industrial IoT settings.}
    \label{fig:applications}
\end{figure}



\subsection{Organization}
The rest of this paper is organized as follows.
Section~\ref{sectionIII} surveys existing literature on zero-trust frameworks and the application of FMs in IoT systems and environments.
Section~\ref{sectionIV} examines critical security threats associated with FMs, including model and data poisoning, model inference attacks, Byzantine faults, backdoor attacks, and challenges in intrusion detection.
Section~\ref{sectionV} explores four foundational principles that form the basis of the ZTFM architecture: LPA, continuous verification, data confidentiality and integrity, and behavioral analytics.
Section~\ref{sectionVI} analyzes the core technical components of ZTFM and their roles in supporting a robust zero-trust architecture.
Section~\ref{sectionVII} identifies open research challenges in the deployment of ZTFM for IoT, emphasizing the necessity of interdisciplinary approaches.
Section~\ref{sectionVIII} concludes the paper.


\section{Status Quo of Zero-trust Frameworks and Foundation Models}
\label{sectionIII}
In this section, we first assess the state-of-the-art (SOTA) in zero-trust frameworks that have primarily focused on privacy protection and data reliability, with limited attention to federated threats, access control, continuous verification, and data integrity. We then examine existing research in FMs relevant to IoT systems and applications.

\subsection{Zero-Trust Frameworks}
Zero-trust frameworks can enhance cybersecurity by continuously verifying access and limiting implicit trust, thereby effectively protecting user privacy against unauthorized exposure. These frameworks can improve data reliability through rigorous authentication and strict access controls so that only trustworthy entities interact with sensitive information.

In~\cite{10138335}, a research agenda was presented to improve security in the Metaverse through a zero-trust continuous authentication framework. The privacy issues associated with implementing continuous authentication in social VR were examined based on a foundational element of the Metaverse. An FL-based adaptive authentication framework that utilizes multimodal biometric data was developed, which can explore biometric authentication for continuous verification in VR. 
FL introduces a privacy-preserving approach that allows collaborative ML across distributed devices while maintaining data confidentiality~\cite{Hu2024OFDMA,Javeed2024adhoc, yuan2023Amplitude,Wei2023Personalized}. However, existing FL protocols remain susceptible to both internal and external adversaries, posing risks to data privacy and system integrity~\cite{Lyu2024tnn}. 

Beyond developing robust global models, it is crucial to design FL frameworks that offer strong privacy guarantees and resilience against various adversarial threats.
Traditional cryptographic protocols, e.g., zero-knowledge proofs and garbled circuits, provide potential solutions for secure computations on private data, however, their scalability remains a major obstacle in large-scale FL systems~\cite{Jere2021ieeesp}. To address this issue, alternative approaches in ZTFM, e.g., LPA, continuous verification, data
confidentiality and integrity, and behavioral analytics, could be explored to enhance the integrity and reliability of user-reported metrics while maintaining privacy.

In~\cite{10288074}, an intelligent connected vehicle system behavior paradigm was built on a zero-trust framework to enhance security and reliability in information perception, communication, and control within a vehicular platoon. The framework can mitigate interference from complex behaviors, information exchange, network topology, and environmental factors.
The authors of~\cite{10764723} explored the transformative impact of AI/ML, blockchain, quantum computing, and cloud/edge technologies on the development and effectiveness of zero-trust architectures. Although these technologies can contribute to advanced trust evaluation and adaptive access control in zero-trust models, it is still difficult to ensure continuous verification and least-privilege access across hybrid and multi-cloud environments.

An analysis of the transition from traditional perimeter-based security to the zero-trust framework was given in~\cite{weinberg2024zerotrustimplementationemerging}. The impact of emerging technologies, such as AI and quantum computing, was explored on zero-trust policies and deployment strategies. In particular, ML in zero-trust was examined, showcasing its ability to enhance security through pattern analysis, anomaly detection, and threat prediction, enabling real-time decision-making.

Complementing these perspectives, Mao et al.~\cite{Mao2023survey6Gedge} provided a comprehensive survey of security and privacy challenges in 6G network edge environments, emphasizing the intersection of Zero-Trust principles with edge computing, AI, and network slicing. The survey identifies key threats, such as resource-constrained adversaries, dynamic trust bootstrapping, and distributed data leakage. Their analysis highlights that while Zero-Trust concepts provide a strong foundation for 6G edge security, practical deployments must contend with unique trade-offs between latency, privacy, and scalability.

\subsection{Foundation Models for Internet of Things}

FM can be leveraged to enhance IoT by providing powerful, generalizable AI capabilities that improve real-time decision-making and automate complex tasks across diverse applications. Their ability to learn from vast amounts of heterogeneous IoT data enables adaptive, scalable, and efficient deployments, greatly advancing the intelligence and responsiveness of IoT systems. 
The key features of FMs that make them applicable to IoT systems include:
\begin{itemize}
    \item Multimodal integration, which aims to fuse and jointly process multimodal data (e.g., images, sensor readings, textual metadata), enhancing IoT systems' situational awareness and context comprehension.
    \item Real-time decision-making, which supports fast inference and real-time responsiveness, essential for latency-sensitive IoT applications, such as healthcare monitoring, industrial automation, and autonomous systems.
    \item Adaptability, where FMs adapt to evolving environmental conditions and dynamic IoT data distributions through minimal additional training.
    \item Representation learning, which extracts meaningful and generalized patterns from noisy, sparse, or multimodal IoT data, improving accuracy and reliability.
\end{itemize}
Pipeline parallelism, data parallelism, and multi-modal learning can be employed to advance the sustainable development of FM in the 6G era~\cite{10558825}. In pipeline parallelism, adapting activation and gradient compression along communication resource allocation helps mitigate communication bottlenecks caused by unstable wireless links.

Network FMs can be designed to capture the distinct characteristics of network data~\cite{guthula2025netfoundfoundationmodelnetwork}. In particular, a network FM incorporates a multi-modal embedding layer to identify cross-modal dependencies between different packet fields for building data representation. 
In~\cite{xue2024leveragingfoundationmodelszeroshot}, a zero-shot IoT sensing was developed with an FM text encoder, which can align IoT data embeddings with semantic embeddings. To enhance the extraction of semantic embeddings, the underlying physics of IoT sensor signals was used in a cross-attention mechanism that integrates a learnable soft prompt, optimized on training data, with an auxiliary hard prompt encoding domain-specific knowledge. 

The authors of~\cite{baris2025foundationmodelscpsiotopportunities} surveyed the potential of FM and large language models (LLMs) in Cyber-Physical Systems (CPS) and the IoT by addressing challenges within the perception, cognition, and communication. Different from traditional task-specific ML models, which face limitations due to data annotation needs and sensor heterogeneity, FM can provide a task-agnostic and self-supervised learning framework that enhances adaptability. Despite their promise, effectively integrating FM and LLMs into CPS-IoT requires moving beyond simplistic adaptations from natural language processing (NLP) and computer vision. 
Implicit neural representations, which encode signals or objects using neural networks, have gained attention as a continuous and memory-efficient alternative to traditional discrete representations. 

The authors of~\cite{gu2025foundationmodelssecretlyunderstand} analyzed leveraging FM to enhance hypernetworks for generalizable implicit neural representation tasks. It confirms that FM can improve hypernetwork performance across labeled and hidden classes, demonstrating adaptability and efficiency in various IoT scenarios.
In addition, training large FM can rely on model parallelism in a decentralized setting over a heterogeneous network~\cite{NEURIPS2022_a37d615b}. In particular, a scheduling algorithm can be designed that distributes computational tasklets across decentralized GPU devices connected via a slow, heterogeneous network. To optimize resource allocation, a formal cost model with an evolutionary algorithm can be used to determine the task distribution strategy that enhances training efficiency.

In~\cite{10599304}, a wireless vision was studied for designing FM tailored to the unique demands of next-generation 6G systems, which aims to enable the AI-native networks. Different from existing NLP-based FM, the proposed framework advocates for the development of large multi-modal models with three core capabilities, namely, processing multi-modal sensing data, grounding physical symbol representations in real-world wireless systems through causal reasoning, and retrieval-augmented generation.
In~\cite{10478189}, a training and serving vision was presented for designing FM, in the aspects of networking, storage, and computing. Parallel training strategies with GPU memory optimization and communication optimization techniques can be conducted, so that each strategy is applied for unique application scenarios. FM can be developed to improve service performance with advanced batch processing, sparse acceleration, and multi-model inference.

\begin{table*}[htb]
\centering
\caption{Related surveys with key applications, technical features, and limitations}
\label{table_surveys}
\begin{tabular}{|p{3cm}|p{4.5cm}|p{4.5cm}|p{4cm}|}
\hline
& Applications examples & Technical features & Limitations \\ \hline

\textbf{Zero-Trust Authentication}~\cite{10138335,10288074,Jere2021ieeesp,10764723,weinberg2024zerotrustimplementationemerging} 
& 
Designed for continuous verification in social VR or vehicular platoons. 
& 
Strengthens security through adaptive access control and reduces reliance on perimeter-based defenses. 
& 
Can be challenging to scale across hybrid environments and raises concerns about privacy and algorithmic bias. 
\\ \hline

\textbf{Privacy-Preserving FL}~\cite{Hu2024OFDMA,Javeed2024adhoc, yuan2023Amplitude,Lyu2024tnn,Wei2023Personalized} 
& 
Enables distributed ML while maintaining data on local devices. 
& 
Protects sensitive user data and supports collaborative model training. 
& 
Involves additional computational overhead and may be vulnerable to internal or external adversaries. 
\\ \hline

\textbf{Foundation Models for IoT and CPS}~\cite{guthula2025netfoundfoundationmodelnetwork,baris2025foundationmodelscpsiotopportunities,xue2024leveragingfoundationmodelszeroshot} 
& 
Provides multi-modal sensing in complex IoT or CPS scenarios. 
& 
Offers a task-agnostic, self-supervised learning framework that enhances adaptability across devices. 
& 
Requires domain-specific innovation and can be difficult to integrate seamlessly into heterogeneous systems. 
\\ \hline

\textbf{Network and Wireless Foundation Models}~\cite{10558825,gu2025foundationmodelssecretlyunderstand,10599304} 
& 
Targeted for 6G networks, incorporating pipeline or data parallelism. 
& 
Improves communication efficiency and real-time data processing in unstable wireless links. 
& 
Implementation complexity is high, and dynamic adaptation remains a significant challenge. 
\\ \hline

\textbf{Decentralized Training of Large Models}~\cite{NEURIPS2022_a37d615b,10478189} 
& 
Uses model parallelism and scheduling for distributed GPU tasks. 
& 
Increases scalability and resource utilization, speeding up training in heterogeneous networks. 
& 
Suffers from high communication overhead, and network heterogeneity can degrade performance. 
\\ \hline
\end{tabular}
\end{table*}

{\color{black}
\subsection{Comparison with Existing Surveys}



Different from the existing surveys, which separately focus on zero-trust authentication frameworks for access control in niche domains (e.g., vehicular networks, metaverse environments) and privacy-preserving federated learning mechanisms (e.g., local model updates, gradient obfuscation), this survey contributes a comprehensive view of ZTFMs that bridges these two lines of research. We delineate the design of ZTFM through four core principles, including Least Privilege Access, Continuous Verification, Data Confidentiality and Integrity, and Behavioral Analytics. We also analyze how they mitigate key attack vectors, such as model poisoning, inference attacks, and insider threats. Moreover, we synthesize technical enablers, including blockchain identity management, TEEs, and federated zero-knowledge proofs (ZKPs), which are not covered holistically in earlier works.

Notably, this survey does not treat zero-trust architecture and FL as isolated paradigms, but instead presents ZTFM as a convergent security framework for future AI-native IoT ecosystems. To the best of our knowledge, it is the first to offer a system-level perspective on how FMs can serve as both targets of protection and active agents of trust enforcement under zero-trust assumptions.
}

\section{Foundation Models and Security Challenges}
\label{sectionIV}



FMs are large-scale, pre-trained ML models that can be adapted to specific tasks with minimal additional training \cite{bommasani2021opportunities}. 
While these models serve as powerful tools for building AI-driven IoT systems, their deployment in decentralized, collaborative environments introduces unique security vulnerabilities. These vulnerabilities stem from adversarial manipulations at various stages of model training and inference, impacting both model integrity and user privacy~\cite{hu2022membership,Shan2023Preserving,Nan2024WWW}. Key security challenges include model and data poisoning, model inference attacks, Byzantine failures, backdoor attacks, and intrusion detection challenges~\cite{Wang2023Adversarial}.


\subsection{Model and Data Poisoning} 
Adversaries employ data poisoning and model poisoning as primary adversarial strategies against FMs, aiming to insert malicious content into the training pipeline to undermine the target model~\cite{Tian2023csur,Yichen2024COMST}. 

\subsubsection{Data Poisoning Attacks}
Data poisoning attacks occur at the data level, where adversaries manipulate the local training data to corrupt the model's learning process. A common strategy is to inject maliciously crafted samples, often mislabeled or containing imperceptible perturbations, into the local datasets of compromised clients. These samples are carefully designed to introduce backdoors or bias the global model’s decision boundaries~\cite{Tian2023csur, Nuding2022codaspy, Jagielski2018sp}.
For example, attackers may insert inputs that associate a specific pattern (e.g., a pixel patch in an image or a phrase in text) with an incorrect target label. Once the global model incorporates updates from poisoned clients, it begins to misclassify inputs containing that pattern, effectively embedding an attack trigger~\cite{Nuding2022codaspy}. These attacks are especially dangerous in FL due to limited visibility into individual client data and the lack of centralized oversight.
Moreover, data poisoning is often stealthy and adaptive. Poisoned data can be sparse, making it hard to detect during aggregation, particularly in non-i.i.d. data environments typical of IoT deployments.


\subsubsection{Model Poisoning Attacks}
\begin{figure}
    \centering
    \includegraphics[width=1\linewidth]{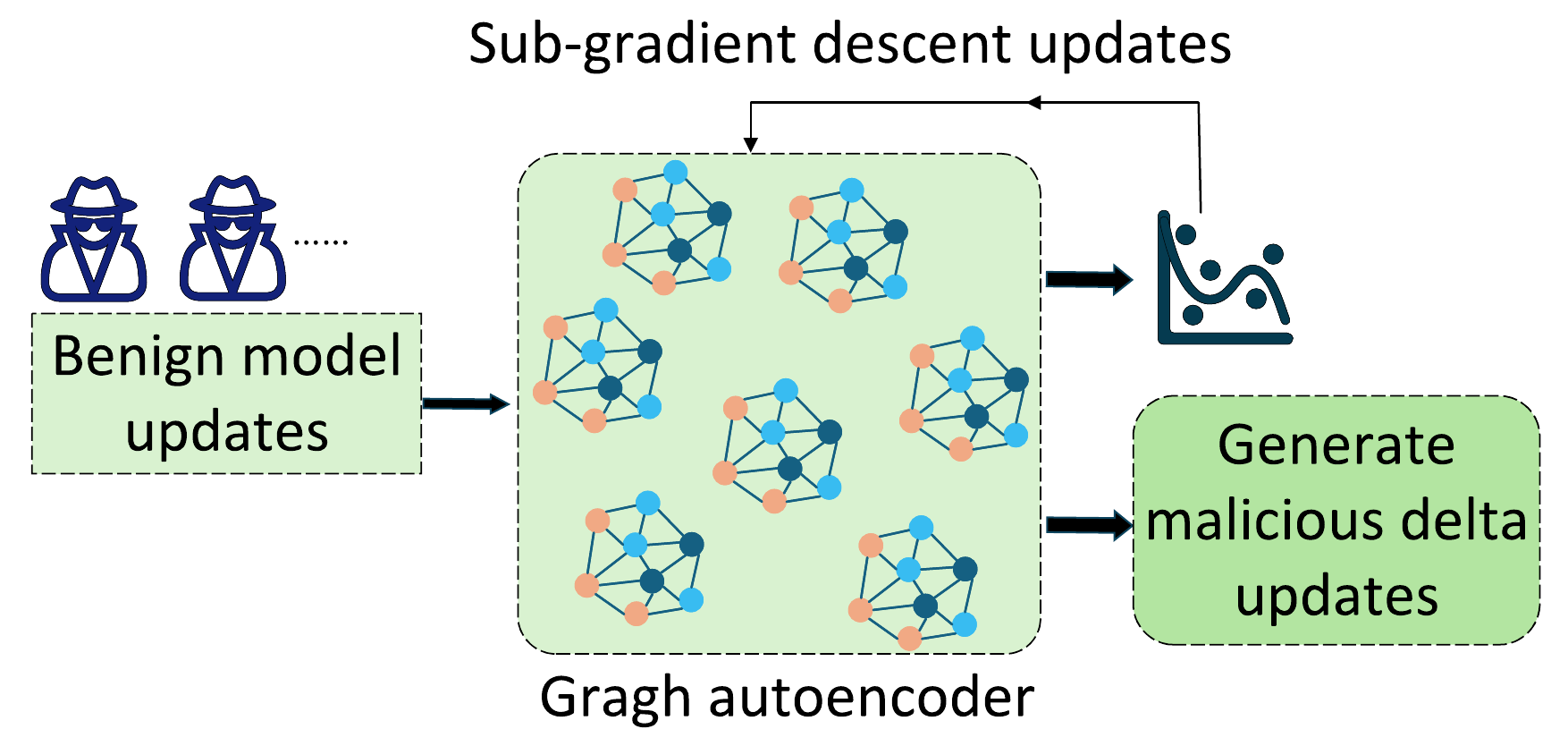}
    \caption{An illustration of an adversarial variational graph autoencoder-enabled model poisoning attack~\cite{Li2025tnn}.}
    \label{fig:gea}
\end{figure}

Model poisoning attacks take place at the model update level. Instead of modifying data, the attacker directly manipulates model parameters or gradients before submitting them to the server. The goal is to inject malicious behavior into the global model or to maximize divergence and disrupt the convergence of the training process.
In~\cite{Li2025tnn}, the authors introduced a sophisticated threat model where an attacker employs an adversarial variational graph autoencoder to infer structural relationships among benign local models, which is depicted in Fig.~\ref{fig:gea}. These relationships are then adversarially modified to generate malicious model updates that still appear statistically similar to benign ones. Notably, this attack operates without requiring access to private data, making it particularly relevant to black-box FL settings such as IoT edge deployments and FL fine-tuning, where access to raw training data is restricted.

Another strategy described in~\cite{Li2024TIFS} enables attackers to eavesdrop on the global model and benign client updates to reconstruct the internal graph structure linking model parameters and data features. The attacker then strategically perturbs these correlations to degrade model performance or introduce targeted behaviors.
This threat is especially concerning for large-scale FMs, where complex feature relationships may be exploited without direct access to local datasets.
In~\cite{Cao2022cvpr}, the focus shifts to malicious user injection, where adversaries inject fraudulent clients that submit poisoned updates aligned with a predefined backdoor objective.
This method is scalable and well-suited for FL environments, such as IoT networks, where participant authentication may be limited. Such vulnerabilities are particularly critical when deploying FMs at the edge, where a few compromised users can influence the model's downstream behavior.

\subsection{Membership Inference Attacks}
Model inference attacks in FL exploit access to shared model updates or intermediate parameters to infer sensitive information such as user attributes, class distributions, or even raw training data~\cite{hu2022membership}. Because FL operates in a decentralized setting, malicious users can execute these attacks without full access to the global model, amplifying the attack surface and complicating privacy preservation.

In~\cite{Gao2023tdsc}, the authors evaluated various secure aggregation protocols and demonstrated that these mechanisms do not fully protect user data. They introduced a differential selection attack combined with de-noising schemes, which allows a malicious actor to infer multi-label classification results from IoT node data, effectively breaching privacy under common FL settings.
A poisoning-assisted inference attack was proposed in~\cite{Wang2023tdsc}, where the attacker leverages benign model updates to extract sensitive feature information. By using a binary attack model, adversaries can identify data patterns not meant to be revealed. Furthermore, a targeted poisoning strategy was introduced, allowing attackers to manipulate training labels and shift the global model's decision boundaries. This manipulation inadvertently causes benign clients to expose additional private feature information.

Inference attacks present serious privacy risks in collaborative learning frameworks such as split learning and FL. In~\cite{Fu2022uss}, both passive and active inference attacks were proposed. The passive attack uses semi-supervised learning on auxiliary data to infer labels, while the active variant manipulates the training process to increase the global model's reliance on the attacker's sub-model, improving inference accuracy.
A broader evaluation in~\cite{liu2022ml} showed that data complexity affects inference attack success, with a trade-off observed between model stealing and inference attack effectiveness. Using architectures like ResNet18 and datasets such as CelebA and Fashion-MNIST, the study highlighted how model behavior varies with data characteristics.
In~\cite{liu2022membership}, membership inference attacks were explored based on prediction sensitivity. By observing how predictions change under small input perturbations, attackers can determine whether a sample was part of the training set, even without knowledge of the model or data, posing a significant privacy threat.


In~\cite{ye2022enhanced}, a hypothesis testing framework was proposed to improve membership inference attacks. The framework uses reference models to boost the true positive rate while keeping the false positive rate under control. The authors also analyzed attacker uncertainty, demonstrating that their method can narrow down uncertainty to a single-bit secret, whether or not a specific data point was part of the training data.
However, this approach relies on well-calibrated reference models, which may be unavailable in practice.
Future extension to black-box settings with limited or unreliable reference models would be desirable.
The authors of~\cite{luo2021feature} explored feature inference attacks at the model prediction stage under a strong adversarial assumption, where the attacker only has access to the model and its outputs.
The attackers could infer private feature values by analyzing model outputs across various architectures. However, its effectiveness relies on having ample prediction samples and may weaken under regularization or dynamic model updates. Future work could explore adaptive defenses or noise-aware attack strategies suited for real-world settings.

\subsection{Byzantine Failure Attacks}
Byzantine failure attacks in FM occur when malicious or faulty users share incorrect or adversarial updates, disrupting the training model's convergence and accuracy~\cite{wu2021tolerating}. These attacks can take various forms, such as random noise injection, adversarial data pollution, software bugs, network asynchrony, or biases in local datasets, making it challenging for aggregation mechanisms to distinguish between benign and malicious contributions~\cite{distler2021byzantine}.

Existing FMs that were considered resilient to Byzantine failures remain susceptible to targeted local model poisoning attacks~\cite{Fang2020uss}. By manipulating the training models from compromised IoT devices, an attacker can significantly degrade the performance of the training model, steering it in a direction opposite to its intended optimization path. While certain defenses adapted from poisoning countermeasures offer partial protection, their effectiveness varies depending on the attack scenario. Existing defense mechanisms against Byzantine attacks were examined, and a vulnerability in FM was argued in~\cite{Shi2022trustcom}. Since the server relies solely on user-reported dataset sizes for weighting updates, without verification due to privacy constraints, malicious IoT devices can manipulate their declared dataset sizes to gain undue influence. Two misreporting strategies were studied, namely, attackers with small datasets falsely claiming to have similar-sized datasets as benign IoT devices, and attackers with comparable datasets inflating their sizes to disproportionately impact the aggregation process.

In~\cite{Blanchard2017nips}, the authors established that existing linear combination methods for aggregating IoT updates cannot withstand a single Byzantine device. A single compromised device can manipulate the FM into selecting an arbitrary model update, potentially with excessive magnitude or a misleading direction. A Byzantine resilience strategy was developed, which can outline sufficient conditions for an aggregation rule to tolerate multiple Byzantine devices.
An FM scheme that simultaneously conducts privacy preservation and resilience against Byzantine failure attacks was presented~\cite{Dong2024tdsc}. The approach employs three-party computation to implement an aggregation method while maintaining the confidentiality of local training models. To enhance efficiency, the scheme includes a maliciously secure top-$k$ protocol with reduced communication overhead and an optimized secure shuffling protocol, which is essential for the secure top-$k$ mechanism.

Both works in~\cite{Blanchard2017nips} and~\cite{Dong2024tdsc} have improved the security of federated systems, but a zero-trust approach requires stricter assumptions, treating all clients and intermediaries as untrusted by default. In this sense, the three-party computation and secure protocols developed in~\cite{Dong2024tdsc} are more in line with zero-trust principles by minimizing reliance on any single party and limiting information exposure. To fully align with zero-trust architecture, these methods could be extended with secure authentication, trusted initialization, and real-time client behavior monitoring.




\subsection{Backdoor Attacks}


Backdoor attacks refer to a type of adversarial attack where an attacker inserts a hidden, malicious trigger into an ML model during its training phase, especially for the FM. The trigger is designed to make the model behave in a specific, undesirable way when exposed to certain inputs, which are typically controlled by the attacker. However, the model's overall performance on normal data remains unaffected, making the attack hard to detect during regular use.

Nguyen et al. \cite{Nguyen2024eaai} study the security risks of backdoor attacks in FL, where malicious participants can secretly insert harmful behaviors into shared models. They review different attack methods and propose various defense strategies, such as anomaly detection and robust aggregation, to make FL more secure.
Wang et al.~\cite{Wang2020nips} explore how backdoor attacks can be inserted into FL systems by targeting the tail distributions of data, which consist of rare or less-represented data points. The authors show that even small changes to the data in these tails can allow attackers to introduce malicious behavior into the model without affecting its overall performance. This demonstrates that FL systems are vulnerable to such backdoor attacks, particularly when the data is imbalanced.

Gong et al.~\cite{Gong2022network} examine coordinated backdoor attacks in FL, where multiple attackers collaborate to insert triggers into the global model. The key insight is that these triggers can be model-dependent, meaning they exploit specific vulnerabilities in the model’s architecture. This makes the backdoor attack more efficient and harder to detect. The paper highlights the challenges in defending against such coordinated attacks in FL systems.
The authors in \cite{Rieger2022ndss} propose a DeepSight that examines the internal structure and outputs of neural network updates to identify and filter out these malicious contributions. It aims to enhance model security without negatively impacting performance on legitimate data.

\subsection{Adversarial Attacks on Intrusion Detection Systems}


\begin{figure}
    \centering
    \includegraphics[width=1\linewidth]{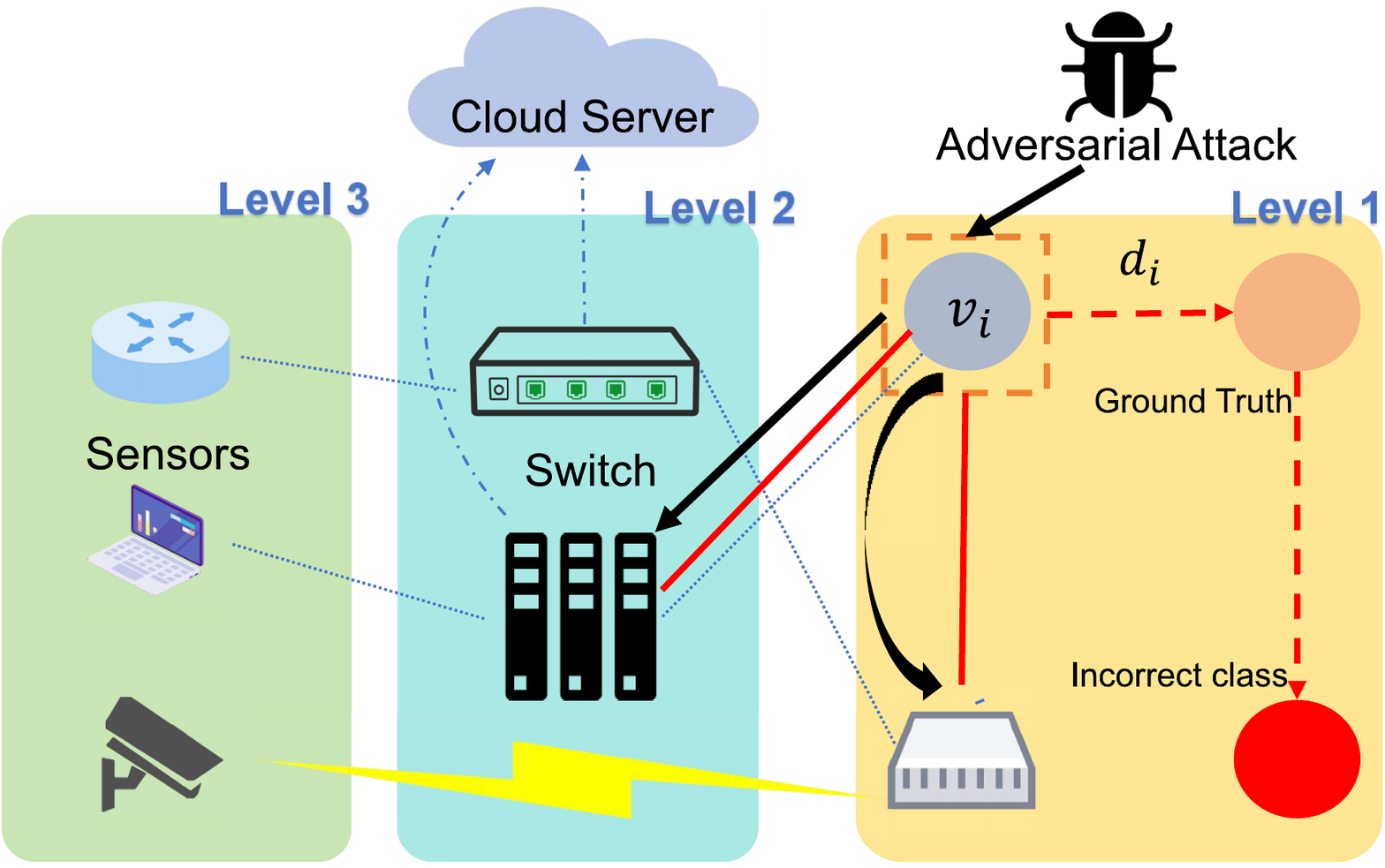}
    \caption{An illustration of hierarchical black-box adversarial attacks on GNN-based NIDS in IoT networks~\cite{Zhou2022iotj}. Here, adversaries exploit multi-level infrastructure and graph feature perturbations to induce misclassifications while bypassing intrusion detection systems.}
    \label{fig:applications}
\end{figure}

Adversarial attacks on Intrusion Detection Systems (IDS) are becoming a bigger issue in cybersecurity. In these attacks, hackers change the way network traffic looks so they can slip past ML systems that are supposed to catch them. They might hide harmful content or make their activity look normal to fool the system. Interestingly, these tricks are very similar to the ones used to fool FLs. 
In both cases, attackers take advantage of how sensitive these models are to small input changes. These changes can cause the model to make mistakes without raising alarms. 
This shows that both IDS and FMs share a common weakness. They can be fooled by well-crafted inputs because they rely heavily on data patterns. 

Zhou et al.~\cite{Zhou2022iotj} investigate the vulnerabilities of graph neural network (GNN)-based IDS used in IoT networks to hierarchical adversarial attacks. The authors present a novel attack framework that targets the structural properties of the IoT network's graph, exploiting the relationships between devices to manipulate the IDS’s detection capabilities. By introducing adversarial perturbations in the graph structure, the attackers can cause the GNN to misclassify malicious activities, thereby undermining the security of the IoT network. 
Dai et al. \cite{dai2018adversarial} introduce adversarial attacks that manipulate graph structures of GNNs and propose a reinforcement learning-based attack strategy that modifies graph structures by adding or removing edges to mislead GNN models in node and graph classification tasks. Moreover, they introduce alternative gradient-based and genetic algorithm attacks for different attack scenarios, including cases with and without access to model gradients. 

These challenges highlight the limitations of traditional FM frameworks, which rely on implicit trust among participants. Addressing these vulnerabilities requires a paradigm shift toward a Z, where no participant or device is inherently trusted, and all interactions are rigorously verified.

\subsection{Defense Strategies Against Adversarial Attacks}
Addressing the vulnerabilities in FM requires a combination of robust defense mechanisms, including anomaly detection, secure aggregation, adversarial filtering, and privacy-preserving techniques. Below, we categorize and discuss key defense strategies that have been proposed to counteract various adversarial threats in FM.

\subsubsection{Defending Against Data Poisoning Attacks}
In~\cite{Cao2022tifs}, an ensemble-based FL framework was proposed. Users are divided into groups, each training a local model. A majority voting scheme is applied during inference to determine the final prediction, reducing the influence of any single poisoned model on the overall outcome. 
The RSim-FL framework~\cite{Chen24www} enhances FL security by using representational similarity analysis. It compares global and local model representations to form a consistency set and applies $K$-means clustering to identify and isolate adversarial users based on representational deviations. This method is applicable to FMs, where consistent representations are important, and supports zero-trust by validating clients based on behavior rather than assumptions of trust.
Building on this, the study in~\cite{Chen2023tifs} introduced a privacy-preserving, hierarchical aggregation defense suited for IoT environments, where edge IoT nodes perform synchronous aggregation under the coordination of a leader node. 
Encrypted poisoned gradients are detected during this process, offering scalability and robustness in heterogeneous settings. The framework can support zero-trust through encrypted communication and multi-level validation, making it suitable for FM-based FL in resource-constrained, decentralized systems.

\begin{figure}
    \centering
    \includegraphics[width=1\linewidth]{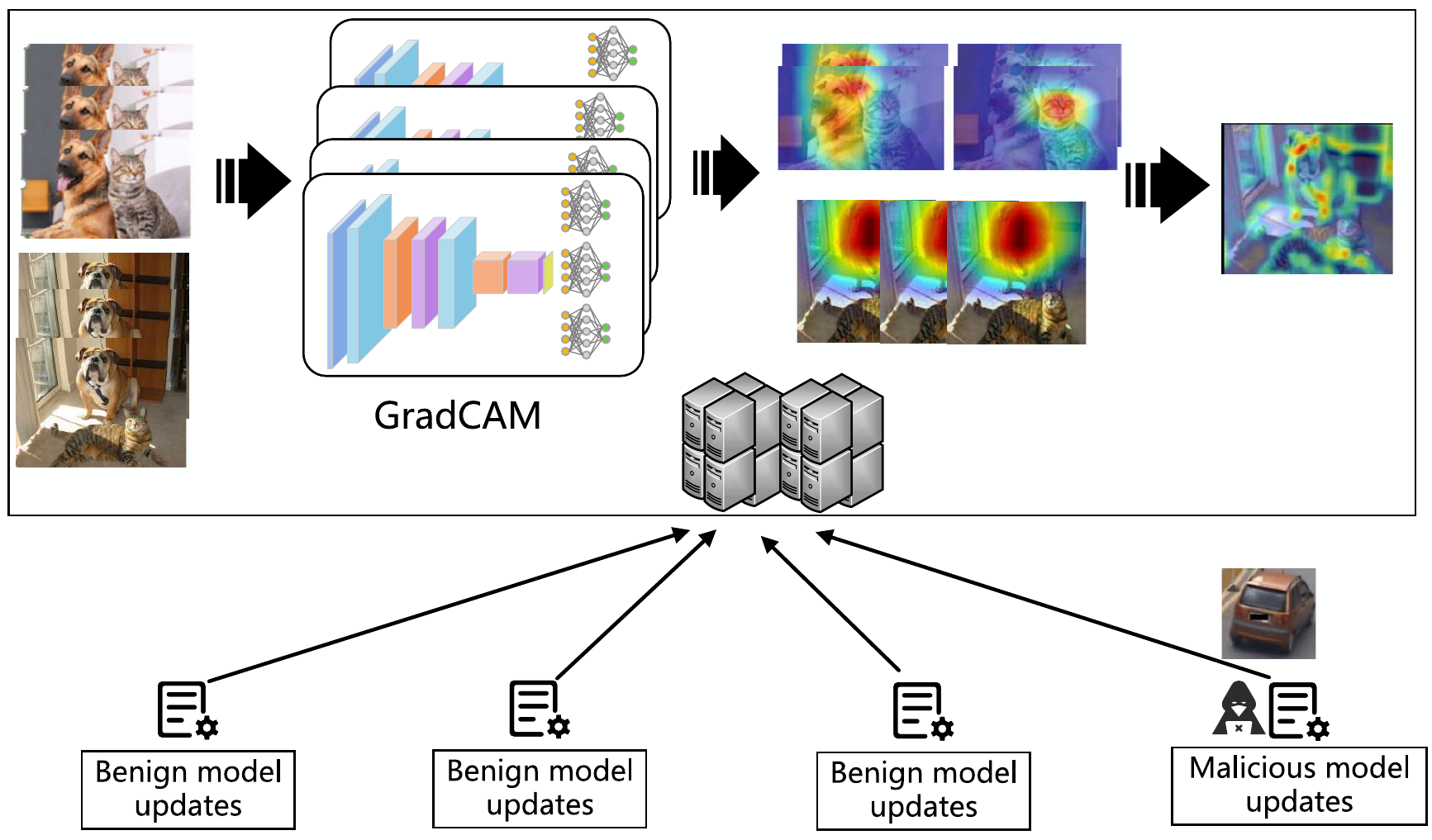}
    \caption{An illustration of the GradCAM-based defense mechanism against model poisoning attacks developed in FMs~\cite{Zheng2024www}.}
    \label{fig:gradcam}
\end{figure}

\subsubsection{Defending Against Model Poisoning Attacks}
Model poisoning attacks manipulate model updates to inject adversarial behaviors into the global FM. To mitigate these threats, recent works have explored advanced detection and aggregation strategies in IoT. 
Zheng et al.~\cite{Zheng2024www} proposed a defense mechanism against model poisoning attacks in FM, which integrates gradient-weighted class activation mapping (GradCAM) and an autoencoder to enhance detection effectiveness beyond traditional Euclidean distance-based methods. As shown in Fig.~\ref{fig:gradcam}, a heat map can be generated by GradCAM for each uploaded local model update, converting it into a lower-dimensional visual representation. This transformation highlights hidden features within the heat maps, improving the ability to detect anomalous patterns and identify malicious local models accurately. 
To enhance security while preserving privacy, a two-trapdoor homomorphic encryption approach was proposed in~\cite{Ma2022tif}. 
In particular, a Byzantine-resilient aggregation incorporating cosine similarity was designed to evaluate the distance between encrypted gradients, aiming to identify encrypted malicious gradients.

Zhang et al.~\cite{Zhang2022kdd} proposed an approach where the FM server employs the Cauchy Mean Value Theorem to predict each user's model updates based on historical trends. 
To assess the consistency of these updates, the Euclidean distance between the predicted and received model updates can be computed for each user. In this case, a suspicious score is assigned to the user, which is adjusted in each iteration to track potential anomalies. 
In~\cite{Chow2021tpsisa}, a targeted perception poisoning attack was developed on FM for object detection, where malicious users inject perception-poisoned local model updates into the federated training process. To mitigate such threats, a spatial signature analysis was studied as a defense mechanism, which differentiates between benign and poisoned model updates to remove adversarial influence and protect the integrity of FM.

\begin{table*}[h!]
\centering
    \caption{Summary of attacks and defense strategies on FMs}
    \label{tab:attacks}
\begin{tabular}{|p{3.0cm}|p{5.25cm}|p{5.25cm}|p{3.0cm}|}
\hline
\textbf{Attack Types} & \textbf{Attack Description} & \textbf{Defense Strategy}  & \textbf{Limitations of Current Defense Strategies} \\
\hline
Data Poisoning Attacks~\cite{Tian2023csur, Nuding2022codaspy, Jagielski2018sp} & Data poisoning attacks occur at
the data level, where adversaries manipulate the local training
data to corrupt the model’s learning process. & 
\begin{itemize}[noitemsep,topsep=0pt,leftmargin=*] 
\item Ensemble FL with Majority Voting~\cite{Cao2022tifs} 
\item RSim-FL with Clustering~\cite{Chen24www}
\item Hierarchical IoT Defense~\cite{Chen2023tifs}
\end{itemize}
& Limited generalizability across dynamic IoT environments; lacks real-time adaptability. 
\\
\hline
Model Poisoning Attacks~\cite{Cao2022tifs, Chen24www, Chen2023tifs, Li2025tnn, Li2024TIFS, Cao2022cvpr} & Data poisoning attacks occur at the model update level, where adversaries manipulate the training process by injecting malicious model updates to degrade the model's performance. & 
\begin{itemize}[noitemsep,topsep=0pt,leftmargin=*] 
\item Feature-Based Anomaly Detection~\cite{Zheng2024www} 
\item Privacy-Preserving Gradient Analysis~\cite{Ma2022tif}
\item Historical Behavior-Based Detection~\cite{Zhang2022kdd}
\item Spatial signature analysis~\cite{Chow2021tpsisa} 
\end{itemize}
& Vulnerable to adaptive strategies that mimic benign behavior; high false negatives. 
\\
\hline
Membership Inference Attacks~\cite{hu2022membership, Gao2023tdsc, Wang2023tdsc, Fu2022uss, liu2022ml, liu2022membership, ye2022enhanced, luo2021feature} & The attacker exploits a trained model’s outputs to infer sensitive information about its training data or parameters. For example, membership inference can reveal if a specific record was in the training set, or model extraction can steal the model. &
\begin{itemize}[noitemsep,topsep=0pt,leftmargin=*] 
\item Adversarial Feature Masking~\cite{shi2022membership} 
\end{itemize}
& Often model-specific and insufficient against adaptive inference; lacks scalability. \\
\hline
Byzantine Failure Attacks~\cite{wu2021tolerating, distler2021byzantine, Fang2020uss, Shi2022trustcom, Blanchard2017nips, Dong2024tdsc} & In distributed or FL, some participants (Byzantine nodes) behave maliciously or unpredictably, sending incorrect or adversarial model updates that corrupt the global model’s training process. & 
\begin{itemize}[noitemsep,topsep=0pt,leftmargin=*] 
\item Divide-and-Conquer Aggregation Algorithm~\cite{Shejwalkar2021ndss}
\end{itemize}
& Ineffective in large-scale or heterogeneous networks; reactive rather than preventive. \\
\hline
Backdoor Attacks~\cite{Nguyen2024eaai, Wang2020nips, Gong2022network, Rieger2022ndss} & The adversary injects a hidden “trigger” pattern into the training data so that the model performs normally on standard inputs but produces an attacker-chosen output when the trigger is present, effectively embedding a backdoor. & \begin{itemize}[noitemsep,topsep=0pt,leftmargin=*] 
\item Adversarial Update Evaluation~\cite{Nguyen2022uss}
\item Certifiable Robustness Against Backdoors~\cite{Xie2021icml}
\item Root-of-Trust Model Verification~\cite{Cao2021ndss} 
\item Trust-Based Scoring of Model Updates~\cite{Huang2024IS} 
\end{itemize}
& Detection methods may require retraining and assume known trigger patterns; impractical for resource-limited IoT devices. \\
\hline
Adversarial Attacks on IDS~\cite{Zhou2022iotj, dai2018adversarial} & The attacker crafts small perturbations to input data at inference time to cause the model to misclassify it (an evasion attack). This is often used to evade security systems (e.g., making malicious network traffic appear benign to an intrusion detection model). & 
\begin{itemize}[noitemsep,topsep=0pt,leftmargin=*] 
\item Adversarial training (augmenting training with adversarial examples) is a primary defense to improve model robustness~\cite{debicha2023advbot}.
\item Input preprocessing or anomaly detection can be applied to identify and reject adversarial inputs~\cite{venturi2023arganids}.
\end{itemize}
& Resource-intensive; poor generalization to novel attacks; brittle under adaptive adversaries. \\
\hline
\end{tabular}
\end{table*}

\subsubsection{Defending Against Model Inference Attacks}

A model inference attack was designed to attack user authentication of FM in 5G and IoT systems~\cite{shi2022membership}. In particular, the model's input consists of received power and phase shift, enabling the attacker to determine whether specific signals were part of the classifier's training data. To execute the attack, the attacker can collect signals and classification results through spectrum observation, construct a surrogate classifier, and apply an inference attack to infer whether a received signal corresponds to one used in the service provider's training dataset.

\subsubsection{Defending Against Byzantine Failure Attacks}
A divide-and-conquer aggregation algorithm was developed to defend against Byzantine failure attacks in FM~\cite{Shejwalkar2021ndss}. Inspired by defenses against poisoning attacks, their divide-and-conquer aggregation identifies and mitigates malicious updates by detecting significant deviations in update space. The algorithm can compute the principal component of the updates, calculate their projections along this direction, and discard a fixed fraction of updates with the largest projections to reduce the impact of adversarial manipulations.

\subsubsection{Defending Against Backdoor Attacks}
Backdoor attacks allow attackers to insert hidden triggers that activate malicious behaviors under specific conditions while maintaining normal performance on regular IoT data.
Nguyen et al. \cite{Nguyen2022uss} propose FLAME, a defense method that mitigates backdoor attacks by evaluating model updates against strategically designed test inputs, ensuring that harmful modifications are identified and filtered out. Unlike traditional defenses, FLAME was designed for the decentralized nature of FL, offering a practical and scalable solution without compromising data privacy.

Xie et al. \cite{Xie2021icml} propose Certifiably Robust FL (CRFL), which applies randomized smoothing techniques to enhance model resilience. Even if an attack successfully poisons a model, the CRFL ensures robustness by enforcing mathematical guarantees that limit backdoor effectiveness.
Cao et al. \cite{Cao2021ndss} proposed FLTrust, a defense mechanism that establishes a ``root of trust'' by maintaining a small, clean dataset at the central server to evaluate updates from all IoT devices and filter out suspicious ones. This approach ensures that FL remains secure and robust against adversarial manipulations, preventing attackers from degrading model performance while maintaining high accuracy.

Gong et al.~\cite{Gong2022network} examine how attackers secretly manipulate FL models by injecting hidden malicious behaviors. They classify these attacks into data poisoning (tampering with training data) and model poisoning (altering model updates), and accordingly categorize defense mechanisms, aiming to provide a structured understanding of existing strategies.  
To address backdoor threats in FL, Huang et al.~\cite{Huang2024IS} introduced Suprte, a trust evaluation mechanism that assigns trust scores to participating devices based on their historical behaviors.
This reduces the influence of suspicious updates and prevents attackers from injecting harmful changes. It is necessary because FL allows multiple IoT devices to train models collaboratively without sharing data, making it vulnerable to hidden attacks that can be difficult to detect using traditional security methods.

\subsubsection{Adversarial Defense to IDS}
An Adv-Bot was proposed in \cite{debicha2023advbot}, which is a framework designed to generate realistic adversarial botnet attacks to bypass network IDS. The authors evaluate various attack strategies and their impact on network IDS performance, showing that adversarial samples can effectively reduce detection accuracy. The study highlights the importance of robust defenses against adversarial attacks in cybersecurity.
Venturi et al. \cite{venturi2023arganids} introduced ARGANIDS, a network IDS leveraging an adversarially regularized graph autoencoder (ARGA) for detecting network anomalies. By incorporating adversarial training, the model improves robustness against evasion attacks and enhances anomaly detection performance. The authors demonstrate that ARGANIDS outperforms traditional network IDS techniques in accuracy and resilience against adversarial modifications.



\subsection{Lessons Learned}

While existing defense methods offer some protection, they incur critical limitations. Techniques like adversarial training, input filtering, or anomaly detection rely on model-specific configurations and are vulnerable to adaptive adversaries that evolve beyond fixed defense strategies. These defenses can struggle to scale across different environments and devices in FM-based IoT systems, 
as FMs, equipped with a general-purpose architecture, are susceptible to subtle and transferable adversarial perturbations that can affect various downstream tasks. 
To this end, the following lessons can be learned:

\begin{itemize}
    \item \textit{Model-agnostic defenses are essential.}
    Relying on hard-coded protections or assumptions about a model's structure or input distribution can leave systems exposed to unseen threats.
    
    \item \textit{Scalability and adaptability must be prioritized.} 
    Defenses must function reliably across different system configurations, data types, and threat models in large IoT networks or when using FMs in different applications.

    \item \textit{Attack resilience must be continuous and proactive.} 
    Static defense is insufficient against adaptive adversaries. Systems should integrate real-time behavioral monitoring, dynamic response mechanisms, and layered protections.
\end{itemize}
    By enforcing strict identity verification, access controls, and ongoing trust evaluation for every device, regardless of location or role, zero-trust can shift the paradigm from perimeter-based defense to comprehensive internal verification. This is useful in FM-empowered IoT networks, where devices frequently connect and disconnect, and decisions must remain verifiable and resilient to unpredictable inputs.




\section{Core Principles of ZTFMs}
\label{sectionV}
ZTFMs combine the expressive power of FMs with the security guarantees of zero-trust architectures, offering a promising pathway for secure, privacy-preserving intelligence in IoT environments. Unlike traditional security paradigms that rely on static trust boundaries, ZTFMs embed continuous verification, fine-grained access control, and secure data sharing into the AI model lifecycle. 
This section elaborates on the four foundational principles underpinning ZTFM, including \textit{LPA}, \textit{Continuous Verification}, \textit{Data Confidentiality and Integrity}, and \textit{Behavioral Analytics}, emphasizing their realization, benefits, and domain-specific challenges.

\begin{figure}
    \centering
    \includegraphics[width=1\linewidth]{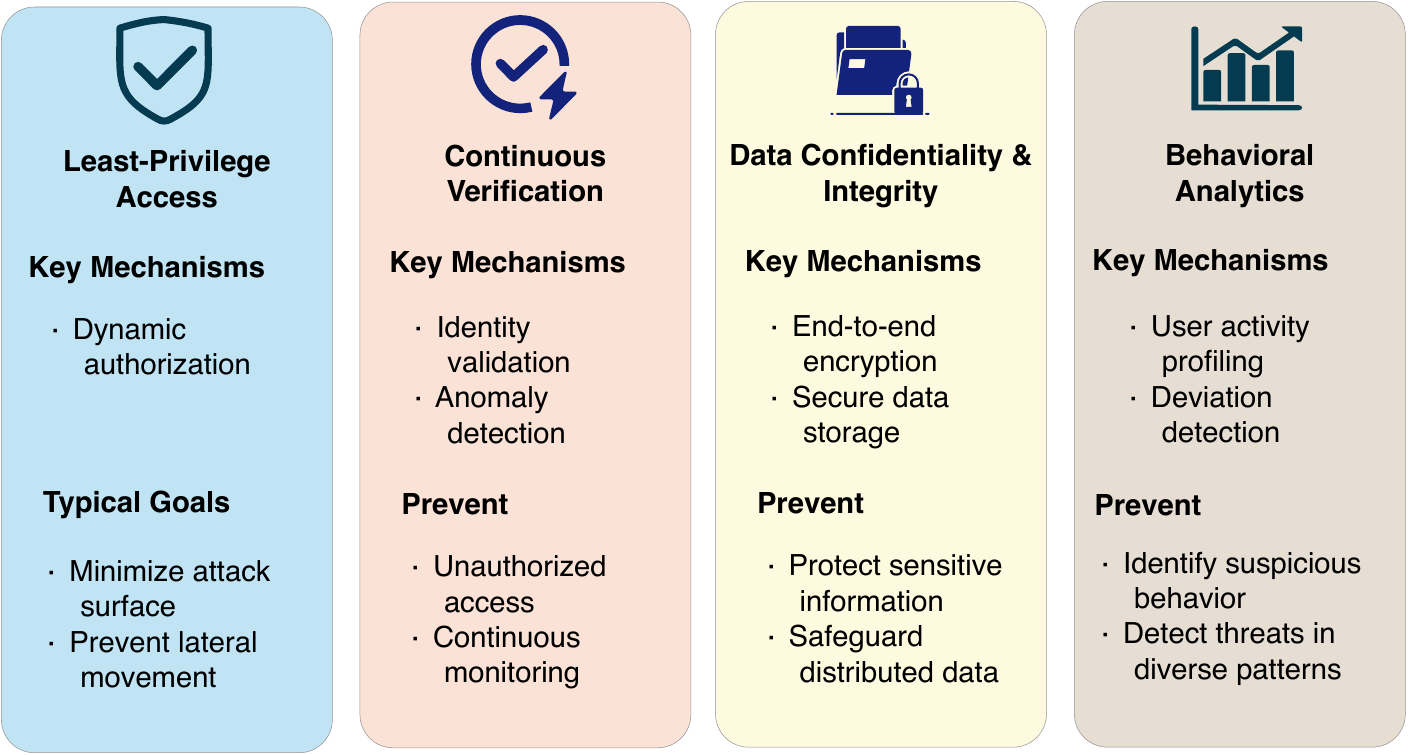}
    \caption{{\color{black}Core principles of ZTFM in IoT environments. Each principle -- Least Privilege Access, Continuous Verification, Data Confidentiality \& Integrity, and Behavioral Analytics -- maps to specific mechanisms and goals designed to enhance security posture in distributed, heterogeneous IoT systems.}}
    \label{fig:corep}
\end{figure}

\begin{table*}[ht]
    \centering
    \renewcommand{\arraystretch}{1.3}
    \caption{Core Principles of ZTFM in IoT and their Targeted Attack Mitigations}
    \label{tab:ztfm_principles}
    \begin{tabular}{|p{2cm}|p{3cm}|p{4.5cm}|p{7cm}|} \hline
        \textbf{Principle} & \textbf{Targeted Attack Type(s)} & \textbf{Advantages} & \textbf{Challenges in IoT} \\
        \hline
        
        \textbf{LPA} &
        Lateral movement, privilege escalation, backdoor persistence & 
        \begin{itemize}
            \item Minimizes attack surface, reducing security risks~\cite{Tuyishime2024ecai2}.
            \item Prevents lateral movement in compromised networks~\cite{Azad2024iot}.
            \item Enables fine-grained, context-aware access control~\cite{kumar2023application}.
        \end{itemize} & 
        \begin{itemize}
            \item \textbf{Complex policy enforcement:} High heterogeneity in IoT makes uniform policy enforcement challenging~\cite{Azad2024iot}.
            \item \textbf{Scalability issues:} Large-scale IoT networks require dynamic access control updates~\cite{kumar2023application}.
            \item \textbf{Lack of standardization:} Many LPA mechanisms are designed for IT environments rather than IoT~\cite{Tuyishime2024ecai2}.
        \end{itemize} \\ 
        \hline
    
        \textbf{Continuous Verification} &
        Identity spoofing, session hijacking, adversarial inference, credential theft & 
        \begin{itemize}
            \item Enables real-time authentication and dynamic revocation based on contextual trust~\cite{dimitrakos2020trust}.
            \item Adapts security policies based on risk assessments~\cite{joshi2024emerging}.
            \item Reduces credential-based attacks with AI-driven anomaly detection~\cite{dong2023securing}.
        \end{itemize} & 
        \begin{itemize}
            \item \textbf{High computational and storage costs:} Continuous verification processes large amounts of IoT data~\cite{dong2023securing}.
            \item \textbf{Latency concerns:} Real-time verification may slow down time-sensitive IoT applications~\cite{joshi2024emerging}.
            \item \textbf{Resource limitations:} Persistent monitoring burdens constrained devices~\cite{dimitrakos2020trust}.
            \item \textbf{Policy fragmentation:} Aligning context-aware trust policies remains complex~\cite{dimitrakos2020trust}.
        \end{itemize} \\ 
        \hline
    
        \textbf{Data Confidentiality \& Integrity} &
        Membership inference, model extraction, data tampering, poisoning attacks & 
        \begin{itemize}
            \item End-to-end encryption protects sensitive IoT data~\cite{choi2023smc}.
            \item Privacy-preserving computations using HE and SMPC~\cite{mohan2024securing}.
            \item Supports compliance in regulated sectors~\cite{zhang2024zero}.
        \end{itemize} & 
        \begin{itemize}
            \item \textbf{Large data volume:} Encryption at scale imposes latency and compute costs~\cite{mohan2024securing}.
            \item \textbf{Key management complexity:} Hard to secure key lifecycle in dynamic IoT settings~\cite{choi2023smc}.
            \item \textbf{Untrusted sources:} Verifying integrity in adversarial environments is resource-intensive~\cite{zhang2024zero}.
        \end{itemize} \\ 
        \hline
    
        \textbf{Behavioral Analytics} &
        Insider threats, anomaly-based evasion, adversarial IDS bypass & 
        \begin{itemize}
            \item Enhances real-time threat detection through AI-driven anomaly detection~\cite{Garcia2022cn}.
            \item Dynamically adjusts access policies based on behavior~\cite{wang2025zero}.
            \item Detects behavioral drift and compromised endpoints~\cite{ameer2024zta}.
        \end{itemize} & 
        \begin{itemize}
            \item \textbf{False positives:} Misclassifications may trigger unnecessary restrictions~\cite{Garcia2022cn}.
            \item \textbf{Computational overhead:} AI models may be infeasible on low-power devices~\cite{wang2025zero}.
            \item \textbf{Privacy risks:} Behavioral tracking may violate regulatory norms~\cite{ameer2024zta}.
        \end{itemize} \\ 
        \hline
    \end{tabular}
\end{table*}

\subsection{Least Privilege Access (LPA)}
LPA ensures that users, devices, and applications are granted only the minimum permissions necessary to fulfill their roles, thereby minimizing the attack surface and preventing lateral movement in the event of a breach. In ZTFM architectures, LPA is implemented through mechanisms such as Just-In-Time (JIT) access, Role-Based Access Control (RBAC), and dynamic policy enforcement.

Tuyishime et al.~\cite{Tuyishime2024ecai2} addressed the security challenges in remote online laboratories, where VPN-based access models often lead to overprivileged access and increase the risk of lateral movement. To mitigate this, they propose Twingate, a ZTNA-based system that enforces per-session, per-resource access policies, thereby aligning with the least privilege principle. The system incorporates micro-segmentation, device compliance checks, and real-time identity verification to ensure that users only access authorized lab environments during specific time windows.
Their evaluation, based on academic lab simulations, demonstrates a significant reduction in attack vectors, especially credential theft and unauthorized privilege escalation.
Such results are particularly relevant for ZTFM systems, where FMs operate across heterogeneous IoT environments. Enforcing least privilege helps ensure that FM-based APIs, inference pipelines, or model updates are not universally exposed, but are accessible only through verified, scoped, and context-aware requests -- reducing the FM attack surface without sacrificing scalability.

Uttecht et al.~\cite{uttecht2020zero} provided a comprehensive zero-trust reference model tailored to U.S. federal government networks. Their architecture outlines how LPA policies, when combined with endpoint verification and policy enforcement points, significantly restrict adversary movement in high-value environments. They emphasize the need for federated identity systems and network segmentation to enforce least-privilege without sacrificing usability.
Chinamanagonda et al.~\cite{chinamanagonda2022zero} analyze LPA enforcement in cloud-native architectures. They present a framework integrating identity federation, context-aware access, and JIT provisioning, demonstrating how microservices and container orchestration tools like Kubernetes can implement LPA policies dynamically. Their findings are particularly relevant for cloud-hosted IoT platforms that require real-time scalability and access control granularity.

Azad et al.~\cite{Azad2024iot} conducted a comprehensive survey on the implementation of zero-trust Architecture within IoT environments, focusing on how foundational principles such as LPA can be operationalized under constrained computing, energy, and communication settings. The authors identify key issues in IoT deployments -- including device heterogeneity, dynamic network topology, and the absence of centralized control -- which collectively complicate the enforcement of static access control policies. To address this, they proposed a lightweight architectural framework composed of decentralized policy agents and hierarchical trust zones. These agents operate at the edge and perform context-aware access evaluations based on device profiles, operational roles, and current security posture. Their approach includes adaptive access control mechanisms and energy-aware decision heuristics, ensuring that access decisions can be made with minimal computational overhead. Their insights are particularly aligned with ZTFM goals, supporting distributed model training and FL coordination in security-critical, low-power environments.

While LPA limits exposure by controlling who can access what, it must be complemented by mechanisms that ensure entities are continuously verified post-authentication, especially in dynamic IoT contexts.


\subsection{Continuous Verification}

{\color{black}
Continuous Verification is a core principle of ZTFMs, requiring that no user, device, or system component be inherently trusted. Every interaction is subject to ongoing validation, behavioral monitoring, and real-time re-authentication. This requirement is especially critical in IoT environments, where dynamic connectivity, identity spoofing, and untrusted endpoints are prevalent.

To enable adaptive and fine-grained verification, Dimitrakos et al.~\cite{dimitrakos2020trust} proposed a trust-aware continuous authorization model for smart home IoT systems. This work extends traditional Attribute-Based Access Control (ABAC) models by embedding a Trust-Level Evaluation Engine (TLEE) directly into the policy lifecycle. Unlike static policies, the TLEE allows access control decisions to adapt in real time based on mutable attributes such as environmental conditions or user behavior. This is particularly beneficial for ZTFMs that require per-session policy enforcement. Their key insight lies in combining contextual trust evaluation with modular policy re-evaluation, ensuring that authorizations remain valid only as long as trust conditions are met. The prototype achieves sub-10ms re-evaluation latency, showcasing its practicality for constrained IoT devices. However, the system’s dependency on locally maintained context data may limit scalability across federated environments.

To address scalability in multi-organizational IoT systems, Joshi et al.~\cite{joshi2024emerging} introduced a graph-based framework for trust propagation. Unlike rule-based verification, their model constructs a dynamic interaction graph, where user-device-service relationships are encoded as context-weighted edges. This approach supports transitive trust computation and detects anomalies through topological drift in the graph structure. A major strength of this framework is its ability to reason about emergent risk across federated domains, enabling fine-grained access decisions in complex settings such as smart cities and industrial automation. However, its reliance on graph maintenance may pose overhead for highly dynamic or low-bandwidth environments.

While the above systems focus on behavioral trust evolution, Adhikari et al.~\cite{adhikari2024advancing} addressed identity privacy during verification, a critical concern in sensitive IoT domains such as healthcare. They proposed a ZKP-based federated identity protocol, which enables users to prove authorization without disclosing personal identifiers. This method improves over traditional OAuth/OpenID flows by eliminating central identity providers and mitigating metadata leakage risks. It supports compliance-driven IoT use cases governed by HIPAA or GDPR. Their performance analysis shows minimal verification overhead, but integrating ZKP at scale may require hardware-assisted acceleration or simplified cryptographic primitives.

In high-mobility and adversarial scenarios, such as UAV delivery networks, Dong et al.~\cite{dong2023securing} developed a comprehensive continuous verification stack. Their system combines biometric-based MFA, continuous behavioral profiling, and blockchain-backed audit trails to ensure persistent identity assurance. A distinguishing feature is the use of immutable blockchain logs to resist rollback or spoofing attempts, making it ideal for mission-critical deployments. The design also incorporates adaptive access control, where deviations in biometric or behavioral profiles trigger automatic revocation. This work demonstrates how ZTFMs can integrate multi-modal identity streams to maintain verification under adversarial conditions. The trade-off lies in the computational burden of blockchain consensus and biometric matching, which may affect real-time responsiveness unless optimized.

From context-adaptive ABAC and graph-based transitive trust to privacy-preserving authentication and tamper-proof verification chains, these latest studies reflect a shift from static, perimeter-based security to intelligent, context-aware, and decentralized trust enforcement.}

\begin{figure}
    \centering
    \includegraphics[width=1\linewidth]{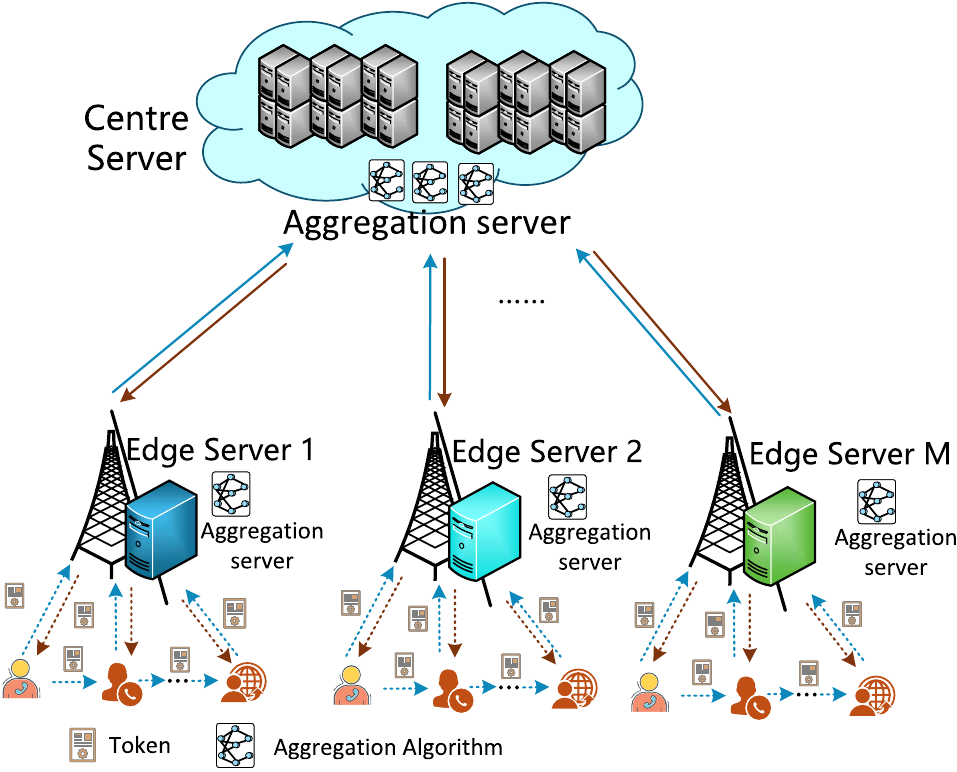}
    \caption{A representative architecture of privacy-preserving FL in edge computing~\cite{choi2023smc}. This hierarchical structure incorporates local aggregation at edge servers and global coordination by a central server, supporting secure multiparty computation and token-based update verification among distributed IoT users.}
    \label{fig:con}
\end{figure}

\begin{figure}
    \centering
    \includegraphics[width=1\linewidth]{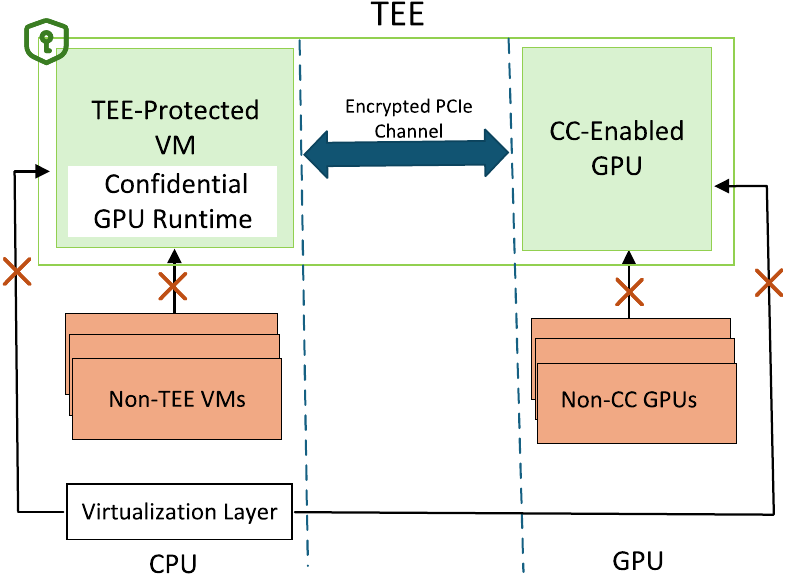}
    \caption{An illustration of a CPU–GPU co-architecture for confidential computing using a TEE, as developed in~\cite{mohan2024securing}. Encrypted PCIe channels and enclave-based isolation prevent unauthorized access from legacy VMs or GPUs, highlighting the role of secure virtualization and hardware-assisted attestation in zero-trust computing environments.}
    \label{fig:con}
\end{figure}

\subsection{Data Confidentiality and Integrity}
Preserving the confidentiality and integrity of data is critical in ZTFM systems, particularly when AI models are collaboratively trained across distributed IoT and edge devices. In ZTFM, this principle is realized through secure computation mechanisms such as homomorphic encryption, SMPC, and TEEs, each offering unique trade-offs in privacy, performance, and scalability.

Choi et al.~\cite{choi2023smc} introduced a scalable SMPC framework tailored for resource-constrained IoT nodes engaged in FL. Their approach partitions training workloads among nodes, ensuring that model gradients remain encrypted and undisclosed during aggregation. They demonstrate resilience against inference attacks even under non-IID data conditions -- a common scenario in IoT -- and evaluate latency-performance trade-offs using real-world health and logistics datasets. Their findings affirm that privacy-preserving collaboration is feasible even under energy and computational limitations.
Zhang et al.~\cite{zhang2024zero} designed a zero-trust security architecture for smart grid telemetry that integrates endpoint authentication, micro-segmentation, and TLS-layer encryption. Their approach not only safeguards the real-time integrity of high-volume telemetry streams but also dynamically evaluates device behavior to adjust trust levels on the fly. This isolation-first model prevents lateral movement and enables fault containment, which is critical in critical infrastructure domains like energy, transportation, and smart manufacturing.

Mohan et al.~\cite{mohan2024securing} explored the feasibility of running large-scale FMs, e.g., BERT and LLaMA, within confidential computing environments using Intel TDX and NVIDIA Hopper. They benchmarked batch inference workloads and demonstrated optimization techniques that reduce the performance penalty introduced by TEEs. Their evaluation shows that while secure inference incurs overhead, batching and parallelism can enable scalable and trustworthy deployment of ZTFMs.
Li et al.~\cite{li2022design} presented the design and formal verification of ARM's confidential compute architecture, which introduces Realms as a trusted execution abstraction. Realms isolate sensitive ZTFM processes from untrusted system software, ensuring data confidentiality even under system-level compromise. Their architecture supports remote attestation, runtime memory encryption, and formal proof of security properties, making it suitable for regulated environments where verifiability and trust assurance are essential.

Across these latest studies, ZTFMs have been designed to support collaborative AI computation while preserving strict confidentiality guarantees. From encrypted model aggregation and secure inference to hardware-enforced isolation, each approach complements zero-trust principles by minimizing data exposure and enabling verifiable trust across dynamic and heterogeneous IoT environments.

\begin{figure}
    \centering
    \includegraphics[width=1\linewidth]{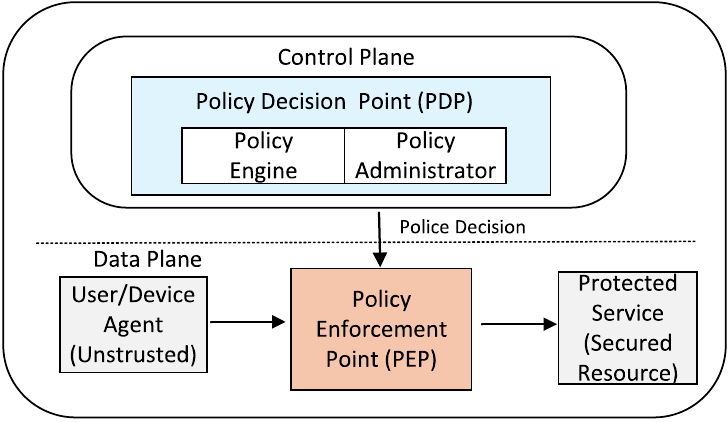}
    \caption{The core structure of the zero-trust architecture proposed in~\cite{wang2025zero}, which delineates the control plane, where the Policy Decision Point (PDP) governs access logic, from the data plane, where the Policy Enforcement Point (PEP) mediates all subject–resource interactions. The separation ensures scalable, policy-driven trust enforcement across dynamic environments.}
    \label{fig:con}
\end{figure}



\subsection{Behavioral Analytics}

{\color{black}
Behavioral Analytics introduces adaptive risk modeling by continuously observing user activity, device state, and environmental context to detect security anomalies and enforce dynamic policy adjustments. In IoT environments -- characterized by frequent device churn, impersonation risks, and context-switching -- this approach is especially important as identity-based controls alone fail to account for behavioral variation.

To operationalize behavior as a dynamic trust signal, Garcia et al.~\cite{Garcia2022cn} proposed SADAC, a Security Attribute-based Dynamic Access Control system. SADAC leverages multivariate statistical process control (MSPC) and behavior-based profiling to adjust access privileges in real time. A key design insight is its modular architecture, enabling seamless integration with enterprise-grade IoT systems. Simulation results show a false positive rate below 3.2\% in insider threat detection scenarios, while maintaining detection latency within 50ms, highlighting both efficiency and robustness for time-sensitive applications.

Moving beyond statistical modeling, Wang et al.~\cite{wang2025zero} introduced a deep learning-based approach, applying Long Short-Term Memory (LSTM) networks to learn temporal access patterns from session metadata, including login intervals, device mobility, and contextual tags. Their evaluation across enterprise and university networks demonstrates a 12–18\% improvement in detection accuracy over baseline heuristics, and precision scores exceeding 90\% in anomaly prediction. The model’s strength lies in its ability to capture long-range dependencies in user behavior, making it suitable for detecting stealthy deviations.

To enable zero-trust enforcement in dynamic environments like healthcare and smart cities, Ameer et al.~\cite{ameer2024zta} proposed a real-time trust scoring framework. The system synthesizes telemetry signals such as device posture, app usage, and geolocation variance, recalculating trust scores every 5–10 seconds. This enables context-sensitive access revocation without relying on static roles or credentials. Their prototype achieved real-time decision-making under 20 ms and demonstrated resilience against impersonation attacks in simulation environments with frequent device switching.

Complementing centralized inference, Kumar et al.~\cite{kumar2023application} adopt a decentralized reputation model by integrating endpoint detection and response (EDR) data with the EigenTrust algorithm. Trust values evolve based on compliance history, audit logs, and peer device evaluations. In industrial IoT testbeds, their system demonstrated a 40\% reduction in incident response time and enabled trust-driven isolation of malicious nodes within 1.5 seconds of anomalous activity detection. This reputation-driven strategy is particularly suited to distributed settings where centralized policy enforcement is infeasible.


In summary, behavioral analytics equips ZTFM with the capacity to perceive, adapt, and respond, not just based on credentials, but on continuous observation and inference. It lays the foundation for proactive threat mitigation in fast-changing IoT ecosystems, bridging context, computation, and trust in a unified zero-trust pipeline.
}

\begin{table*}[ht]
\centering
\renewcommand{\arraystretch}{1.4}
\caption{Mapping Technical Components to Core ZTFM Principles in IoT}
\label{tab:ztfm_mapping}
\begin{tabular}{|p{3cm}|p{3.2cm}|p{3.2cm}|p{3.2cm}|p{3.2cm}|}
\hline
\textbf{Technical Component} & \textbf{LPA} & \textbf{Continuous Verification} & \textbf{Data Confidentiality \& Integrity} & \textbf{Behavioral Analytics} \\
\hline

\textbf{Authentication \& Authorization} &
JIT access control and identity-based RBAC enforce minimal access scope \cite{Tuyishime2024ecai2,chinamanagonda2022zero} & 
Trust-aware continuous authorization and federated identity mechanisms support contextual verification over time \cite{dimitrakos2020trust,adhikari2024advancing} & 
Authentication ensures only authorized nodes access encrypted data streams~\cite{adhikari2024advancing} & 
Contextual identity and access logs serve as input for dynamic behavior scoring \cite{Nagarajan2024tce} \\
\hline

\textbf{Secure Aggregation} &
Limits what each node contributes; prevents excessive exposure of internal models \cite{Hussain2025TCE} & 
Secure model update paths enable verification without revealing raw data \cite{pokhrel2024robust} & 
SMPC and DP protect model updates and gradients during FL \cite{choi2023smc} & 
Anomaly-aware aggregation reduces the impact of malicious updates \cite{pokhrel2024robust} \\
\hline

\textbf{Anomaly Detection} &
Detects policy violations in access logs and restricts permissions dynamically \cite{Garcia2022cn} & 
Continuously assesses behavior of devices/users to verify consistency \cite{wang2025zero} & 
Flags suspicious access to encrypted or sensitive data, enhancing data integrity \cite{Nawshin2024adhoc} & 
Builds user/device profiles to identify drift, insider threats, or outliers in behavior \cite{Javeed2024adhoc} \\
\hline

\textbf{Blockchain} &
Smart contracts enforce decentralized, fine-grained access policies in dynamic networks \cite{Gupta2023comcom} & 
Immutable logs support auditability of verification and access decisions \cite{dong2023securing} & 
Distributed ledgers protect model sharing and prevent data tampering \cite{asante2021distributed} & 
Blockchain anchors behavior logs and device reputation scores \cite{Liu2024cybersec} \\
\hline

\textbf{Trusted Execution Environments (TEEs)} &
Securely enforces policy checks within isolated execution zones \cite{li2022design} & 
Supports attestation and runtime validation of model integrity and updates \cite{vomvas2024establishing} & 
Executes encrypted models and data securely, protecting confidentiality \cite{mohan2024securing} & 
Ensures that analytics models are protected from tampering while processing sensitive behavior data \cite{li2024tees} \\
\hline

\textbf{Encryption \& Secure Communication} &
Encrypts channel-specific access tokens to restrict user scope \cite{gharib2023scc5g} & 
TLS and mTLS encrypt sessions to support session-level continuous verification \cite{rodigari2021performance} & 
Ensures end-to-end encrypted data transmission with PQC and EDAP \cite{tseng2021encrypted} & 
Encrypted traffic metadata may be analyzed to detect behavioral anomalies without revealing content \cite{kim2024anomaly} \\
\hline

\end{tabular}
\end{table*}


\section{Technical Components of ZTFMs} 
\label{sectionVI}


\begin{table*}[ht]
    \centering
    \renewcommand{\arraystretch}{1.3}
    \caption{Technical Components of ZTFM in IoT}
    \label{tab:ztfm_principles}
    \begin{tabular}{|p{5cm}|p{6cm}|p{5cm}|p{4cm}|}
        \hline
        \textbf{Technology} & \textbf{Advantanges} & \textbf{Challenges in IoT} \\ 
        \hline
        Authentication \& Authorization \cite{Hussain2025TCE,Nagarajan2024tce,Gupta2023comcom,Hussain2024wc,dash2024zero} & Robust security via multi-factor, blockchain, AI-driven verification; dynamic continuous validation & Increased complexity; computational overhead; possible latency \\
        \hline
        Secure Aggregation \cite{Hussain2025TCE,pokhrel2024robust,Xie2024wc,10701059,aiello2023secure} & Data confidentiality; protection against malicious updates; decentralized aggregation enhances resilience & Communication overhead; potential accuracy loss due to privacy mechanisms; computational complexity \\
        \hline
        Anomaly Detection \cite{Nawshin2024adhoc,Ramezanpour2022cn,Xie2024wc,Javeed2024adhoc,pokhrel2024robust} & Early threat identification; real-time response; detection of adversarial activities & False positives; resource-intensive; requires continuous monitoring \\
        \hline
        Blockchain \cite{Hussain2025TCE,kim2023blockchain,Sullivan2024iccnc,pooja2024secure,jain2024emerging,Liu2024cybersec} & Immutable auditability; transparency; decentralized trust and security management & High energy and computational overhead; scalability limitations; latency in transaction validation \\
        \hline
        Trusted Execution Environments (TEEs) \cite{li2024tees,Liu2024iotj,vomvas2024establishing} & Strong isolation of sensitive computations; protection against tampering and leakage & Hardware dependency; potential performance overhead; complexity in deployment and maintenance \\
        \hline
        Encryption \& Secure Communication \cite{gharib2023scc5g,tseng2021encrypted,rodigari2021performance} & Ensures data confidentiality; protects against interception; strong cryptographic security & Latency and overhead due to encryption processing; complexity in key management; resource-intensive \\
        \hline       
    \end{tabular}
\end{table*}

ZTFM is underpinned by a suite of integrated technical components that work together to enforce strict security postures across distributed systems. These components are designed to ensure that trust is never assumed, and every interaction is continuously verified and evaluated. From the moment entities request access, through the secure handling and aggregation of data, to the detection of anomalous behaviors and the preservation of data integrity, each element contributes to a robust, end-to-end trust framework. This section provides an in-depth look at the core technical elements of ZTFM and their roles in a resilient zero-trust architecture.

\begin{figure}
    \centering
    \includegraphics[width=1\linewidth]{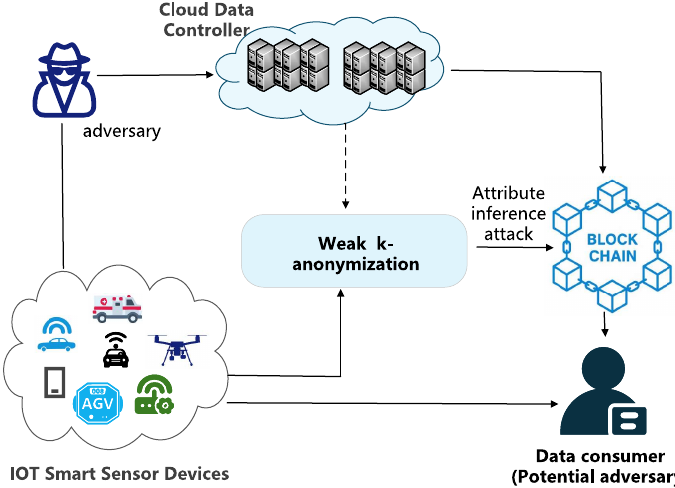}
    \caption{Overview of a potential privacy breach in blockchain-based IoT data sharing~\cite{Hussain2025TCE}. This attacker model illustrates the end-to-end flow of data from IoT devices to adversarial consumers, emphasizing the risks of using weak anonymization and highlighting the need for stronger privacy-preserving mechanisms in Zero-Trust IoT ecosystems.}
    \label{fig:con}
\end{figure}

\begin{figure}
    \centering
    \includegraphics[width=1\linewidth]{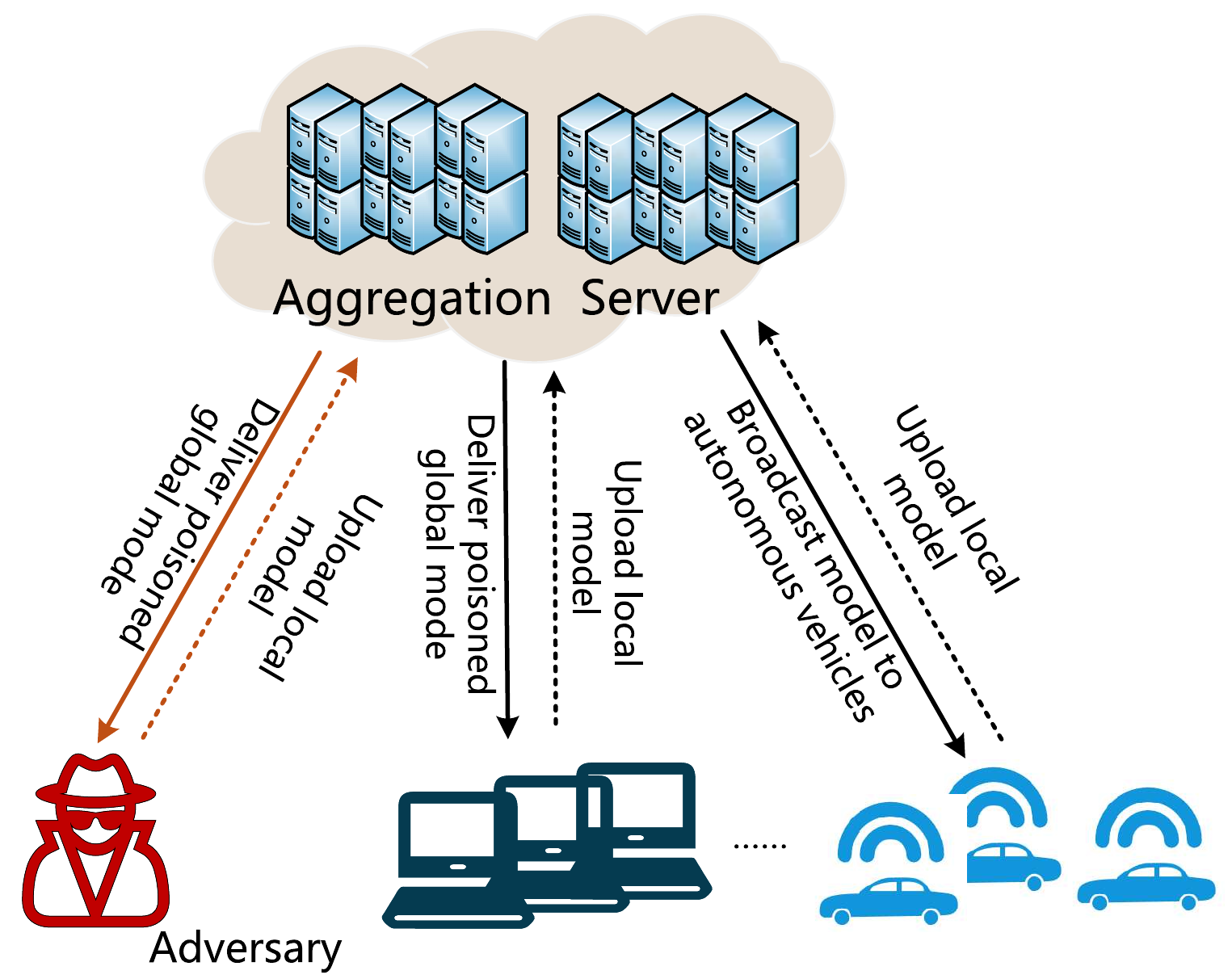}
    \caption{Workflow of an FL poisoning attack presented in~\cite{Hussain2025TCE}. The adversary manipulates the local training process to upload poisoned models. The aggregation server fails to detect the malicious updates and integrates them into the global model, which is then disseminated to all connected devices, compromising system integrity.}
    \label{fig:con}
\end{figure}

\subsection{Authentication and Authorization} 
Robust identity verification is critical for securing FL in IoT, ensuring that only trusted participants join model training~\cite{kairouz2021advances}.
In the context of ZTFM, this prevents unauthorized devices from introducing poisoned updates or launching inference attacks on FMs.
Mechanisms, such as multi-factor authentication, digital certificates, and identity management systems~\cite{Hussain2025TCE}, can help secure participation, while integrating local differential privacy, FL, and blockchain ensures scalable, tamper-resistant verification across IoT environments.
The AIDL-XTS model developed in~\cite{Nagarajan2024tce} demonstrates how AI models (e.g., CNN-BiLSTM) can profile user and device behavior for continuous trust scoring, which aligns with ZTFM's need for adaptive, real-time verification.
Moreover, proxy smart contracts~\cite{Gupta2023comcom} validate transactions before finalization, and can offer a blueprint for securing model updates and access decisions in distributed FMs.

FL-based dynamic access control frameworks~\cite{Hussain2024wc} and continuous verification engines~\cite{edo2024zero,dash2024zero} showcase how zero-trust principles like behavior analysis, micro-segmentation, and contextual access can enhance ZTFM resilience in IoT systems.
Continuous authentication and access control have been shown to improve cyber resilience in large-scale IoT networks~\cite{Liu2024cybersec}, reinforcing dynamic trust management for FMs in adversarial, resource-constrained environments.
These studies highlight that existing zero-trust solutions, though not explicitly designed for FMs, offer mechanisms that ZTFM extends for securing FM-based IoT systems.

\subsection{Secure Aggregation} 
Aggregation protocols ensure confidentiality by securely combining model updates from participants without exposing sensitive data, employing differential privacy to prevent reconstruction attacks while maintaining model accuracy~\cite{Hussain2025TCE}. Blockchain-based FL methods proposed in \cite{pokhrel2024robust} counteract malicious client updates, reinforcing global learning security. Blockchain integrated with dynamic zero-trust FL \cite{Xie2024wc} enhances data privacy and security within industrial IoT environments. Moreover, privacy-preserving aggregation protocols coupled with main-side blockchain architectures further secure consumer IoT data \cite{10701059}.

A framework for zero-trust verification of industrial IoT (IIoT) wireless transmission nodes was developed in \cite{Xie2024wc}, which utilizes FL to achieve zero-trust rule training and terminal model training, while employing blockchain technology for on-chain aggregation and cloud backup of the models. This approach enhances the accuracy and availability of the zero-trust rules while safeguarding the security of IIoT nodes.

\subsection{Anomaly Detection}

Anomaly detection for FMs within a zero-trust architecture differs fundamentally from traditional anomaly detection in IoT systems. Traditional IoT anomaly detection identifies irregular patterns in sensor data, network traffic, or device behavior using predefined rules or lightweight models, often limited to narrow contexts and static trust assumptions. In contrast, anomaly detection for FMs in zero-trust environments operates at multiple abstraction levels, detecting not only data-level anomalies but also adversarial manipulations, distribution shifts, unauthorized model access, or malicious behavior embedded in complex model interactions. This requires continuous verification of both data and model behavior, leveraging behavioral analytics, provenance tracking, and trust scoring. Moreover, zero-trust detection has to account for the higher adaptability of FMs, ensuring model integrity and confidentiality even under stealthy, sophisticated threats that go beyond outliers or threshold violations targeted in legacy IoT anomaly systems.

Advanced algorithms detect and isolate anomalies such as adversarial updates and unexpected communication patterns, safeguarding collaborative learning integrity~\cite{chen2025trustworthy}. The DP-RFECV-FNN framework~\cite{Nawshin2024adhoc} leverages differential privacy and deep learning to classify and prevent unauthorized Android malware in IoT networks. Continuous monitoring combined with AI-driven dynamic trust algorithms ensures real-time risk evaluation and access control in 5G/6G networks~\cite{Ramezanpour2022cn}. Additionally, ML techniques proposed in~\cite{Xie2024wc} detect anomalies in industrial IoT data streams, effectively identifying internal and external threats. CNN and BiLSTM integrated frameworks further enhance anomaly detection by capturing spatial-temporal patterns in evolving cyber threats~\cite{Javeed2024adhoc}. Pokhrel et al.~\cite{pokhrel2024robust} introduced a robust zero-trust architecture integrating blockchain and FL, enhancing anomaly detection and securing decentralized IoT networks. A privacy-preserving AI-driven malware detection framework was proposed in~\cite{Nawshin2024adhoc} for IoT-based medical devices running on Android. Integrating differential privacy and zero-trust security ensures secure, decentralized malware detection, safeguarding sensitive patient data and healthcare network integrity while maintaining high accuracy.

A zero-trust framework for smart grid infrastructures was proposed in~\cite{10758677}, which integrates IT and OT security mechanisms to enhance monitoring and defense against sophisticated cyber threats, such as ransomware. By leveraging EigenGame for data integration and quantum reinforcement learning for malicious behavior detection, the framework strengthens cybersecurity in IIoT-enabled smart grids, ensuring reliable system protection and threat mitigation.

\subsection{Blockchain} 
Blockchains provide immutable transaction records, enhancing transparency and auditability in collaborative AI model training \cite{Hussain2025TCE,wang2019survey,Jiang2024ACMSurvey}. For instance, Kim et al. \cite{kim2023blockchain} demonstrated blockchain applications in securing FL, ensuring transparency in collaborative AI updates. Blockchain-based protocols proposed by Sullivan et al. \cite{Sullivan2024iccnc} securely endorse real-time vehicle trajectory data. Blockchain integration within zero-trust architecture improves transparency, security, and access control for scientific peer review and data sharing~\cite{pooja2024secure}, while Jain et al.~\cite{jain2024emerging} highlighted blockchain’s role in securing healthcare data alongside AI-driven threat detection. Liu et al. \cite{Liu2024cybersec} offered a comprehensive bibliometric analysis, identifying significant blockchain-based trends in zero-trust IoT security research, emphasizing its effectiveness against heterogeneous device environments.

A blockchain and smart contract-based edge-IoT framework was proposed in \cite{Pan2019iotj}, which enforces zero-trust security by managing IoT device behavior through a credit-based resource allocation system, ensuring secure access control, automated policy enforcement, and scalable security in decentralized IoT networks. A blockchain-based middleware for management in IoT was presented in \cite{Samanieg02018iciot}, which uses a novel zero-trust hierarchical mining process that allows validating the infrastructure and transactions at different levels of trust.

\subsection{Trusted Execution Environment} 
Hardware-based TEEs protect sensitive computations and model parameters from tampering or leakage \cite{schneider2022sok}. The applicability of TEEs in decentralized AI systems is explored extensively, highlighting secure computational capabilities \cite{li2024tees, Liu2024iotj}. For example, Vomvas et al. \cite{vomvas2024establishing} proposed a vertical extension termed zero-trust execution for beyond-5G networks, using TEEs to secure execution environments and establish trust in untrusted contexts.
In addition, Aiello et al. \cite{aiello2023secure} examined the secure access service edge framework, integrating SD-WAN, ZTNA, SWG, and CASB, emphasizing identity-driven security, micro-segmentation, and real-time threat intelligence to address network performance and data protection challenges.

\subsection{Encryption and Secure Communication:} 
End-to-end encryption protocols, such as TLS, safeguard data confidentiality and integrity, mitigating interception risks during data transmission \cite{hazra2024data}. Gharib et al. \cite{gharib2023scc5g} introduced SCC5G, a Post-Quantum Cryptography (PQC)-based architecture ensuring encrypted and authenticated 5G mission-critical communications, utilizing CRYSTALS-Kyber and CRYSTALS-Dilithium cryptographic schemes.
Tseng et al.~\cite{tseng2021encrypted} proposed Encrypted Data Processing (EDAP), which employs processor-level encryption to secure data during execution, eliminating implicit trust in cloud platforms and hypervisors. Additionally, Rodigari et al. \cite{rodigari2021performance} assessed mutual TLS (mTLS) in zero-trust networks, confirming its security effectiveness despite moderate computational overheads in multi-cloud deployments.
A zero-trust architecture optimized for industrial environments was proposed in \cite{Zanasi2024adhoc}, integrating micro-segmentation and software-defined networking to enhance security in power grids, transportation systems, and industrial control systems. By enabling dynamic network management, granular access control, and breach containment, the framework strengthens cybersecurity in highly heterogeneous and interconnected industrial networks.

\subsection{Lessons Learned}
Most current solutions have addressed some isolated aspects of zero trust or FMs, and often overlooked the dual role of FMs as both targets and enablers of trust enforcement. ZTFM can potentially bridge this gap by positioning FMs as active agents within the trust pipeline, enabling real-time threat detection, dynamic access control, and adaptive policy enforcement based on contextual understanding.

Effective ZTFM implementations necessitate the holistic integration of multiple technologies, e.g., authentication, anomaly detection, blockchain, TEEs, and encryption, into a unified trust framework. 
While mechanisms like multi-factor authentication, continuous verification, and differential privacy strengthen security, they introduce computational overheads. Designing lightweight and adaptive solutions remains an ongoing challenge in resource-constrained IoT environments.

Last but not least, static access controls are insufficient in dynamic IoT systems. Techniques, such as behavioral analytics and contextual policy adaptation, are crucial for maintaining continuous trust in the presence of evolving threats. While many works focus on standard security goals, robust defenses against adversarial attacks targeting FMs, such as model poisoning, inference leakage, and stealthy backdoors, are still in their early stages. There is an urgent need for ZTFM-specific adversarial defense mechanisms.

\begin{figure}
    \centering
    \includegraphics[width=1\linewidth]{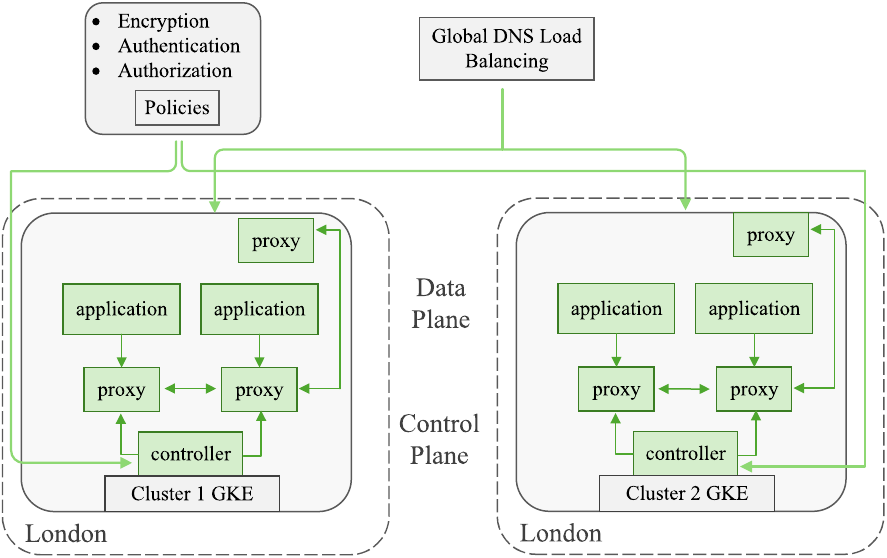}
    \caption{Illustration of the zero trust multi-cloud architecture with distributed policy enforcement via proxies developed in~\cite{rodigari2021performance}, enabling secure communication between services across GKE and EKS clusters.}
    \label{fig:con}
\end{figure}


\section{Open Research Challenges for ZTFM in IoT}
\label{sectionVII}
Despite the progress in ZTFM, several critical research challenges remain unresolved when applying these models to IoT due to the highly distributed and heterogeneous nature of IoT devices, severe resource constraints, scalable and decentralized network conditions, diverse sensitivity and privacy levels of IoT data, and the necessity of lightweight yet robust continuous verification and adaptive security mechanisms across decentralized deployments. 
Based on our study, we identify the following open challenges that demand further investigation:

\subsubsection{Lightweight Cryptographic Primitives for Trust-Aware AI}
Cryptographic methods such as homomorphic encryption, zero-knowledge proofs, and SMPC are core enablers of ZTFM but are resource-intensive for IoT devices. A key challenge lies in developing lightweight cryptographic protocols~\cite{KUMAR202431} that maintain rigorous security guarantees while being computationally feasible for constrained edge nodes. Emerging directions include post-quantum cryptography~\cite{liu2024post} tailored for FL and efficient lattice-based encryption~\cite{ramakrishna2024novel} adapted to non-IID data distributions in IoT. 

\subsubsection{Scalable and Adaptive Trust Reasoning}
As ZTFM systems scale across diverse IoT networks, trust scoring must evolve to account for context, temporal changes, and inter-node variability. One challenge is designing hierarchical, decentralized trust models that incorporate behavioral analytics, reputation systems, and federated signals without incurring excessive synchronization overhead. Graph-based trust propagation~\cite{joshi2024emerging} and dynamic Bayesian belief updating~\cite{zhao2024atypical} are promising but underexplored directions.

\subsubsection{Cross-Domain Interoperability and Policy Federation}
ZTFM deployments across healthcare, manufacturing, and transportation sectors face challenges due to heterogeneity in policies, data formats~\cite{malik2018methodology}, and access requirements~\cite{sicari2015security}. A proposed research direction is to design extensible policy languages and cross-domain security ontologies that can support secure interoperation across trust domains. Additionally, trust federation protocols that preserve local autonomy while enabling global policy compliance are essential.

\subsubsection{Threat-Resilient Federated Training Architectures}
ZTFM must account for adversarial adaptation in collaborative learning pipelines~\cite{nguyen2025zero}. A key challenge is integrating robust FMs protocols with zero-trust guarantees, capable of defending against model poisoning, sybil attacks, and gradient leakage~\cite{elzemity2024privacy}. Future research should explore adversarial training integrated with zero-trust scoring, secure aggregation via threshold cryptography, and active defense using anomaly-triggered retraining.

\subsubsection{Fine-Grained Resource-Aware Security Orchestration}
IoT environments present inherent constraints in bandwidth, memory, and compute~\cite{said2023bandwidth,langguth2018memory}. A pressing challenge is orchestrating ZTFM security enforcement, such as micro-segmentation, continuous verification, and behavioral scoring, in a resource-adaptive manner. This includes dynamic policy offloading, opportunistic security tasks scheduling, and energy-aware trust checkpoints to optimize for security-utility trade-offs \cite{ren2025zero}.

\subsubsection{Auditability and Explainability in Zero-Trust Decisions}
As ZTFM decisions govern sensitive access control and collaboration workflows, the lack of transparent decision paths limits trust and compliance~\cite{crowther2024blending}. We propose integrating explainable AI techniques into ZTFM enforcement modules, enabling audit trails, user-centric justification of denial/approval, and provenance tracking of policy adaptation in distributed environments \cite{li2024making}.

\textbf{Summary.} Collectively, these open challenges highlight the need for multidisciplinary solutions that combine security, cryptography, systems design, and AI to realize scalable, interpretable, and efficient ZTFM deployment in real-world IoT ecosystems.

\section{Conclusion}
\label{sectionVIII}
This paper establishes ZTFMs as a transformative approach to securing AI-driven IoT systems. By embedding zero-trust principles, such as LPA, continuous verification, data confidentiality, and behavioral analytics, into the training and deployment of FMs, ZTFMs provide a principled framework for addressing the unique security and trust challenges of decentralized, heterogeneous IoT environments. The structured synthesis of ZTFMs was presented with formalized foundational principles. A unified technical architecture was proposed, which integrates FL, blockchain-based identity management, micro-segmentation, and TEEs. Our analysis of emerging threats and corresponding defense strategies revealed key limitations in current practices and uncovered several open research directions, including scalable secure orchestration, lightweight multiparty computation, interpretable threat attribution, and AI-driven trust calibration. 

\bibliographystyle{IEEEtran} 
\bibliography{references}

\end{document}